\documentclass[reprint,amsmath,amssymb,aps,floatfix,prr,longbibliography]{revtex4-2}

\usepackage{graphicx}
\usepackage{dcolumn}
\usepackage{bm}
\usepackage{marvosym}
\usepackage{tcolorbox}
\tcbuselibrary{breakable}
\usepackage{mwe}
\usepackage{lipsum}
\usepackage{printlen}
\usepackage[normalem]{ulem}

\usepackage{xcolor}
\usepackage[hidelinks]{hyperref}
\definecolor{fireEngine}{RGB}{193,39,45}
\definecolor{navyBlue}{RGB}{0,0,139}
\definecolor{teal}{RGB}{69,126,136}
\definecolor{caledon}{RGB}{172,234,189}
\definecolor{naplesYellow}{RGB}{246,210,88}

\usepackage{physics}
\usepackage{mathtools}
\usepackage{amsmath}
\usepackage{bbm}
\usepackage[caption=false]{subfig}
\usepackage{float}
\usepackage{empheq}

\usepackage{tikz}
\usetikzlibrary{matrix,calc, arrows.meta, automata, positioning, quotes}

\usepackage{tabularray}
\UseTblrLibrary{booktabs}

\newcommand{\paren}[1]{\left( #1 \right)}
\newcommand{\sbrac}[1]{\left[ #1 \right]}
\newcommand{\curly}[1]{\left\lbrace #1 \right\rbrace}
\newcommand{\angbr}[1]{\left\langle #1 \right\rangle}
\newcommand{\e}[1]{\text{e}^{#1}}
\newcommand{\md}{\mathrm{d}}

\date{\today}

\begin{document}

\title{Global thermodynamic manifold for conservative control of stochastic systems}

\author{Jordan R.\ Sawchuk}
\email{jordan\_sawchuk@sfu.ca}
\author{David A.\ Sivak}
\email{dsivak@sfu.ca}
\affiliation{Department of Physics, Simon Fraser University, Burnaby, British Columbia, Canada V5A1S6}

\date{\today}
\begin{abstract}
    Optimal control of stochastic systems plays a central role in nonequilibrium physics, with applications in the study of biological molecular motors and the design of single-molecule experiments. While exact analytic solutions to optimization problems are rare, under slow driving conditions, the problem can be reformulated geometrically solely in terms of equilibrium properties. In this framework, minimum-work protocols are geodesics on a thermodynamic manifold whose metric is a generalized friction tensor. Here, we introduce a foundation for this friction-tensor formalism for conservatively driven systems on either discrete- or continuous-state spaces. Under complete control of the potential energy, a global thermodynamic manifold (on which points are identified with instantaneous energy landscapes) has as its metric a full-control friction tensor. Arbitrary partial-control friction tensors arise naturally as inherited metrics on submanifolds of this global manifold. We leverage a simple mathematical relationship between system dynamics and the geometry of the global manifold to derive expressions for the generalized friction and linear-response excess work. We show that the friction tensor, usually defined as an infinite-time integral, may be expressed as a product of straightforwardly computed matrices. We furthermore show that the linear-response excess power is decomposable into relaxation-time-scaled projections of the control velocity onto eigenmodes of the rate matrix or Fokker-Planck operator. This intuitive result provides a systematic means of interpreting otherwise-mysterious efficient control strategies, unveiling how multi-parameter optimal control takes advantage of multi-exponential relaxation dynamics to reduce dissipation relative to single-parameter control. We demonstrate the computational and conceptual advances in three illustrative examples.

\end{abstract}

\maketitle
\section{Introduction}
A central question in nonequilibrium thermodynamics concerns the ``price of haste''~\cite{andresenCurrentTrendsFinitetime2011}. For quasistatic (infinitely slow) processes, thermodynamic systems remain at equilibrium, rendering total dissipation independent of driving protocol. However, for processes occurring in finite time, established results from equilibrium thermodynamics no longer apply: instantaneous values of macroscopic observables fundamentally depend on the driving history. For mesoscopic systems where the relevant energy scales are comparable to thermal fluctuations, these quantities are stochastic, exhibiting protocol-dependent distributions~\cite{seifertStochasticThermodynamicsFluctuation2012}.

These factors pose challenges to uncovering universal principles governing systems of theoretical and practical interest like biological molecular motors operating in the nonequilibrium environment of living cells---often with remarkable speed and efficiency~\cite{brownTheoryNonequilibriumFree2020}. Optimal control theory is one framework for navigating the complexity of nonequilibrium statistical physics: given a cost function and boundary conditions, optimization selects a set of privileged paths from the uncountable infinity of thermodynamically distinct paths in control-parameter space~\cite{blaberOptimalControlStochastic2023, guery-odelin_driving_2023}. 

Minimum-work protocols (MWPs) minimize the ensemble average of the excess work, the extra energy required to drive a process in finite time. Although MWPs have been analytically derived for some simple systems~\cite{schmiedlOptimalFinitetimeProcesses2007, gomez-marin_optimal_2008, espositoFiniteTimeThermodynamics2010, aurell_refined_2012, plata_optimal_2019, zhongLimitedcontrolOptimalProtocols2022}, the problem is typically intractable. Reformulating the optimal-control problem as a system of Hamiltonian equations allows for the application of numerical methods for ordinary differential equations~\cite{zhongLimitedcontrolOptimalProtocols2022}; however, computation rapidly becomes prohibitively expensive for large systems, high-dimensional control vectors, or systems driven near a critical point. Ongoing efforts to address these challenges include methods employing machine learning~\cite{yanLearningNonequilibriumControl2022, engelOptimalControlNonequilibrium2023a}.

An alternative approach is to apply suitable approximations that simplify the optimal control problem. Approximations of the excess work based on purely equilibrium information have been derived in the limits of both fast~\cite{blaberStepsMinimizeDissipation2021} and slow~\cite{sivakThermodynamicMetricsOptimal2012, bonancaOptimalDrivingIsothermal2014} driving. The slow-driving approximation gives rise to a formalism within which MWPs may be viewed as geodesics on a thermodynamic manifold. The metric on this manifold is a generalized friction tensor that depends only on the instantaneous value of the control parameters. 

The friction-tensor formalism has been applied in various contexts, including the study of driven barrier crossings as a model of single-molecule force-extension experiments~\cite{sivakThermodynamicGeometryMinimumdissipation2016}, driving the biological rotary motor F$_1$ ATPase~\cite{luceroOptimalControlRotary2019, guptaOptimalControlF$_1$ATPase2022}, folding of DNA hairpins~\cite{tafoyaUsingSystemsEquilibrium2019}, control of protein copy-number distributions~\cite{blaberOptimalControlProtein2020}, bounds on the efficiency of membrane separation of binary gases~\cite{chenGeodesicLowerBound2023}, and nonequilibrium phase transitions~\cite{deffnerKibbleZurekScalingIrreversible2017}. The formalism has also been extended beyond the classical regime to quantum systems~\cite{zulkowskiOptimalControlOverdamped2015}. A re-weighted friction tensor has been introduced for the efficient sampling of free-energy surfaces~\cite{lindahlRiemannMetricApproach2018} with applications to the binding affinity of drugs~\cite{lundborgPathOptimalAlchemistry2023}. Recently it was shown for continuous overdamped dynamics that the thermodynamic length defined in the friction-tensor formalism is equivalent to the $L^2$-Wasserstein distance from optimal-transport theory under the restriction to equilibrium distributions~\cite{zhongLinearResponseEquivalence2024}.

In this paper, we establish a foundational geometric structure for the friction-tensor formalism. We define a non-Riemannian \textit{global thermodynamic manifold} $\mathcal{M}$ whose metric tensor~(\ref{eq:fullconttensordef}) is the generalized friction tensor obtained by assuming arbitrary control over the microstate energies of a given system. We demonstrate how, in the case of conservative control, arbitrary partial-control friction tensors arise naturally as inherited metrics~(\ref{eq:inheritedMetric}) on submanifolds of $\mathcal{M}$. We then derive two decompositions~(\ref{eq:dynamicalDecomposition}, \ref{eq:spectralDecomposition}) of the full-control friction tensor that facilitate numerical and analytic calculations, provide insight into the dynamical origins of dissipation for slowly driven systems, and suggest optimization heuristics. We provide concrete demonstrations of the framework in three illustrative examples.

\section{Conventions and assumptions \label{sec:prelims}}
Our principal results apply to both discrete and continuous state spaces $\Omega$ under a set of assumptions about the nature of the dynamics and control. It will thus be advantageous to establish a unified notation that does not require any assumptions on the cardinality of the state space.  

Summation (for discrete state spaces) and integration (for continuous state spaces) are unified by introducing a measure $\mu$ on $\Omega$: if $g(x)$ is some measurable function on $\Omega$,
\begin{equation}
    \int_{\Omega}\md\mu(x)\,g(x) = \begin{dcases}
        \sum_{x_\nu \in\Omega}g(x_\nu) \ , \ \ &
        \Omega \text{ discrete} \\
        \int_{\Omega}\md x \, g(x) \ , &
        \Omega \text{ continuous}
    \end{dcases} \ .
\end{equation}
Formally, $\mu$ is the \textit{counting measure} for discrete $\Omega$ and the \textit{Lebesgue measure} for continuous $\Omega$~\cite{cheneyAnalysis}, i.e., the notation above indicates either a sum or an integral depending on whether $\Omega$ is discrete or continuous. Then for any state functions $g$ and $h$, we denote by 
\begin{equation} \label{eq:innerproduct}
    \paren{g,h} \equiv \int_{\Omega}\md \mu(x) \, g(x)h(x) \ 
\end{equation}
the Euclidean $\ell^2$ inner product (dot product) for discrete $\Omega$ or the Euclidean $L^2$ inner product for continuous $\Omega$. In particular, 
\begin{equation}
    \angbr{g}_{q} \equiv \paren{g,q} 
\end{equation}
is the average value of $g$ with respect to a distribution $q$ over $\Omega$. 

Throughout, we extend in the straightforward way the definition of the inner product~\eqref{eq:innerproduct} to arguments other than state functions [e.g., $\paren{\bm u, \bm v} = \bm u^{\mathsf{T}}\bm v$ for vectors $\bm u, \bm v \in \mathbb{R}^m$].

We consider ergodic continuous-time Markov processes $X_t$ taking values $x$ in $\Omega$. The time evolution of the probability distribution $p(x,t)$ over $\Omega$ is governed by a generator $\mathcal{G}$, i.e., $\partial p /\partial t = \mathcal{G}p$. External control is accomplished through a vector of control parameters $\bm{u}$ such that $\mathcal{G}_t = \mathcal{G}_{\bm{u}(t)}$. The most general equation of motion for $p(x,t)$ is then
\begin{subequations} \label{eq:MEnFPE}
    \begin{align}
        \frac{\partial p}{\partial t}(x,t) &= \big[\mathcal{G}_{t}p(t)\big](x) \\ 
        &\equiv \int_{\Omega}\md \mu(y) \, \mathcal{G}_t(x,y)p(y,t) \ , \label{eq:MNKernel}
    \end{align}
\end{subequations}
where in \eqref{eq:MNKernel} we define the integral kernel $\mathcal{G}_t(x,y)$ of $\mathcal{G}_t$, i.e., the rate of probability flowing to $x$ from $y$. For discrete $\Omega$, $\mathcal{G}_t$ is a rate matrix $\mathbb{W}_t$, and this is the master equation. For continuous $\Omega$, $\mathcal{G}_t$ is the second-order differential Fokker-Planck operator $\mathcal{L}_t$. The notation $[\mathcal{G}_tp(t)](x)$ here reflects that $\mathcal{G}_t$ acts on the entire function $p(t)$ [just as $\mathbb{W}_t$ acts on the entire vector $\bm{p}(t)$]. 

We further assume that the fixed-parameter dynamics are overdamped and reversible, i.e., the generator $\mathcal{G}_{\bm{u}}$ satisfies detailed balance for all $\bm{u}$:
\begin{equation} \label{eq:detailedBalance}
    \mathcal{G}_{\bm{u}}(y,x)\pi_{\bm{u}}(x) = \mathcal{G}_{\bm{u}}(x,y)\pi_{\bm{u}}(y) \ ,
\end{equation}
where $\pi_{\bm{u}}$ is a distribution satisfying $\mathcal{G}_{\bm{u}}\pi_{\bm{u}} = 0$. Under the assumptions of ergodicity and detailed balance, $\pi_{\bm{u}} \propto \e{-\beta V(\bm{u})}$ is the unique equilibrium distribution for the process at fixed $\bm{u}$, for potential energy $V$ and inverse temperature $\beta \equiv (k_{\rm B}T)^{-1}$. Driving may thus be expressed by a time-dependent potential energy $V(t) = V(\bm{u}(t))$, i.e., control is \textit{conservative}~\cite{remleinOptimalityNonconservativeDriving2021}.

We often write $\angbr{g}_{p(t)}$ as $\angbr{g(t)}$ for the average of a state function with respect to a time-dependent distribution $p(t)$. When the parameters are temporally varied, $\pi(t) \equiv \pi_{\bm{u}(t)}$ is the equilibrium distribution corresponding to the parameter values at time $t$. The mean deviation of a state function $g$ from its instantaneous equilibrium value is $\langle \delta g(t) \rangle \equiv (g, \delta p(t))$ for $\delta p (t) \equiv p(t) - \pi(t)$. Equilibrium (fixed-parameter) time-correlation functions are 
\begin{subequations}
    \begin{align}
        \Sigma_g(t) &\equiv \angbr{\delta g(t) \delta  g(0)}_{\text{eq}} \\
        &= \int_{\Omega}\md \mu(x)\int_{\Omega} \md \mu(y) \, P(x,t|y,0) \pi(y) \, \delta g(x) \delta g(y) \ ,
    \end{align}
\end{subequations}
where $P(x,t|y,0)=\sbrac{\e{t \mathcal{G}}}\!(x,y)$ is the probability that the system is in state $x$ at time $t$ given that it was initialized in state $y$ at time $0$.

With additional and not-too-restrictive assumptions on $V(x,t)$ for continuous systems \footnote{The technical condition (the Bakry-Emery criterion) is that $V$ be $\lambda$-convex~\cite{pavliotis_stochastic_2014}. A quadratic potential in $\mathbb{R}^d$ is an example of a $\lambda$-convex potential.}, these conditions mean that $\mathcal{G}$ has a discrete spectrum of real, non-positive eigenvalues~\cite{pavliotis_stochastic_2014}. 

The right eigenfunctions $\psi_{\alpha}$ and left eigenfunctions $\phi_{\alpha}$ satisfy
\begin{subequations}
    \begin{align}
        \mathcal{G}\psi_{\alpha} &= \lambda_\alpha\psi_{\alpha} \\
        \mathcal{G}^{\dagger}\phi_{\alpha} &= \lambda_{\alpha}\phi_{\alpha} \ . \label{eq:leftEigenfuncts}
    \end{align}
\end{subequations}
Here the adjoint $\mathcal{G}^\dagger$ is defined by $\paren{h, \mathcal{G} g} = \paren{\mathcal{G}^{\dagger}h, g}$. For discrete $\Omega$, the adjoint operator is the transpose $\mathbb{W}^{\mathsf{T}}$ of the rate matrix, and for continuous $\Omega$ it is the backwards Kolmogorov operator $\mathcal{L}^\dagger$. In particular, with $\lambda_0 = 0$, $\psi_0 = \pi$ and $\phi_0 = 1$. The left and right eigenfunctions form a complete dual basis, i.e., they can be normalized such that $(\psi_{\alpha}, \phi_{\beta}) = \delta_{\alpha\beta}$ for Kronecker delta $\delta_{\alpha\beta}$, and
\begin{equation}
    \sum_{\alpha}\psi_{\alpha}(x)\phi_{\alpha}(y) = \delta(x,y) \ ,
\end{equation}
with $\delta(x,y)$ the Dirac delta function when $x$ and $y$ are continuous states and the Kronecker delta function when they are discrete states. Consequently, the fixed-$\bm{u}$ generator admits a spectral decomposition
\begin{equation} \label{eq:generatorSpectralDecomposition}
    \mathcal{G}_t(x,y)=\sum_{\alpha \geq 1}\lambda_{\alpha}(t)\psi_{\alpha}(x,t)\phi_{\alpha}(y,t) \ ,
\end{equation}
such that Eq.~\eqref{eq:MNKernel} can be written as
\begin{equation}
    \frac{\partial p}{\partial t}(x,t) = \sum_{\alpha \geq 1} \lambda_{\alpha}(t)\, \big( \phi_{\alpha}(t), p(t) \big)\, \psi_{\alpha}(x,t) \ ,
\end{equation}  
with $\lambda_{\alpha}$, $\phi_{\alpha}$, and $\psi_{\alpha}$ in general depending on time through the control parameters $\bm{u}$. 

We do not use Einstein summation in this paper due to potential ambiguities caused by expressions such as \eqref{eq:generatorSpectralDecomposition}. However, for objects with clear geometric interpretations, we do distinguish vectors and covectors by upper and lower indices, respectively. For example, control parameters are written as $u^i$, and their conjugate forces are written with a lower index as $f_i$. Latin indices are reserved for parametric control-parameter sets, while Greek indices are used for state-vector indices for discrete systems (e.g., probabilities $p_{\nu} \equiv p(x_\nu)$) and for eigenfunction indices across all systems.

\section{Thermodynamics, control, and geometry \label{sec:frictiontensorformalism}}
Given a set of control parameters $\bm{u}$ and some cost functional $\mathcal{C}$, an optimal control protocol is a curve $\bm{u}^*:\sbrac{0,\tau}\to\mathbb{R}^m$ that minimizes the cost functional subject to some set of constraints. The choice of cost function depends on the context: For example, one may wish to minimize entropy production or the loss of available work, maximize power output or efficiency, or optimize some combination of these~\cite{salamonFiniteTimeOptimizations1981, andresenCurrentTrendsFinitetime2011}. In this paper, we study protocols that minimize the average work subject to a set of holonomic constraints $\bm{V}(t) =\bm{V}(\bm{u}(t))$ that define the control parameters. 

Over a particular realization of a conservative control protocol, the energy $V(X_t ,t)$ of the system at time $t$ depends on time through both the deterministic evolution $V(t)$ of the potential energy and the stochastic dynamics $X_t$. On the ensemble level, these contributions can be separated in the time-derivative of the mean system energy $\angbr{V(t)}_{p(t)} = \paren{V(t), p(t)}$ into the mean power and mean heat flow~\cite{sekimotoLangevinEquationThermodynamics1998a}:
\begin{equation}
    \frac{\md }{\md t}\paren{V(t), p(t)} = \underbrace{\paren{\frac{\md V(t)}{\md t}, p(t)}}_{\text{mean power } \angbr{\mathcal{P}(t)}} + \underbrace{\paren{V(t), \frac{\md p(t)}{\md t}}}_{\text{mean heat flow } \langle\dot{Q}(t)\rangle} \ .
\end{equation}
For a protocol $V(t)$ over a time interval $[0,\tau]$, the average work (taken over many instantiations of the control protocol) is obtained by integrating $\angbr{\mathcal{P}(t)}$ over the protocol:
\begin{equation} \label{eq:excessWork}
    \angbr{\mathcal{W}} = \int_0^{\tau}\md t \paren{\frac{\md V(t)}{\md t}, \, p (t)} \ .
\end{equation}
It is convenient to instead minimize the mean \textit{excess} work (or irreversible work), defined as the mean work minus the quasistatic work  $\angbr{\mathcal{W}}^{(\text{QS})} = \lim_{\tau \to \infty}\angbr{\mathcal{W}}$, obtained by taking $p(t) \to \pi(t)$ in \eqref{eq:excessWork}. One may verify that the quasistatic work so defined is independent of $\tau$ and is equal to the difference in free energy between the initial and final states.
\begin{equation} \label{eq:exactexcessworkV}
    \angbr{\mathcal{W}_{\text{ex}}} = \int_0^{\tau}\md t \, \paren{ \frac{\md V}{\md t}, \delta p (t)} \ .
\end{equation}
This can equivalently be written as
\begin{equation} \label{eq:exactexcesswork}
    \langle \mathcal{W}_{\text{ex}}\rangle = - \int_0^\tau \md t \paren{\frac{\md\bm{u}}{\md t},\left\langle \delta \bm{f}(t)\right\rangle} \ 
\end{equation}
in terms of control parameters $\bm{u}$ and mean conjugate-force deviations
\begin{equation} \label{eq:conjugateForces}
    \angbr{\delta \bm{f}(t)}  \equiv -\paren{\angbr{\nabla_{\bm{u}}V}_{p(t)} - \angbr{\nabla_{\bm{u}}V}_{\pi(t)} } \ .
\end{equation}

\begin{figure}
    \centering
    \includegraphics[width=\linewidth]{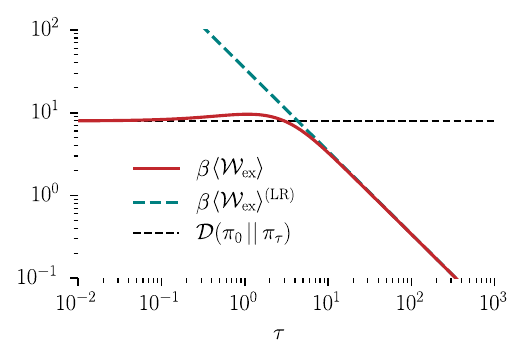}
    \caption{Convergence of the exact excess work $\beta\langle\mathcal{W}_{\text{ex}}\rangle$ (solid red curve) to the linear-response approximation $\beta\langle\mathcal{W}_{\text{ex}}\rangle^{(\text{LR})}$ (dashed teal line) as the protocol duration $\tau$ becomes large, for the $(h,J_{\text{NN}}, J_{\text{NNN}}, K, L)$ protocol described in Sec.~\ref{ssec:spinSystems}. As $\tau \to 0$, the excess work asymptotically approaches the relative entropy $D(\pi_0 \| \pi_{\tau})$ between the initial and final equilibrium distributions~\cite{blaberStepsMinimizeDissipation2021} (horizontal dashed black line).}
    \label{fig:workConvergence}
\end{figure} 

We assume that the controlled system is prepared at time $t = -\infty$ with some initial value of the control parameters $\bm{u}_0$ so that $p(0) = \pi_{\bm{u}_0}$. If the system is driven such that it remains close to equilibrium throughout the protocol, dynamic linear-response (LR) theory~\cite{Chandler1987a} approximates the average deviation of the conjugate forces $\bm{f}$ to the parameters $\bm{u}$ as
\begin{equation}
    \left\langle \delta \bm{f}(t) \right\rangle \approx \beta\int_{-\infty}^t \md t^\prime \, \frac{\md \Sigma_{\bm{f}}}{\md t}\left[\bm{u}(t^\prime) - \bm{u}(t)\right] \ ,
\end{equation} 
for equilibrium time-correlation matrix $\Sigma_{\bm{f}}(t) = \angbr{\delta \bm{f}(t)\delta \bm{f}(0)^{\mathsf{T}}}_{\text{eq}}$. The derivative $\md \Sigma_{\bm{f}}/\md t$ characterizes the system response at time $t$ to a perturbation of the control parameters at time $0$. 

If driving is slow compared to the system's relaxation timescales, the conjugate-force deviation reduces to
\begin{equation} \label{eq:LRrelpartial}
        \left\langle \delta \bm{f}(t) \right\rangle \approx -\tilde{\zeta}(\bm{u})\frac{\md \bm{u}}{\md t} \ .  
\end{equation}
for generalized friction tensor~\cite{sivakThermodynamicMetricsOptimal2012}
\begin{equation} \label{eq:frictensdef} 
  \tilde{\zeta}(\bm{u}) \equiv \beta\int_0^\infty \md t'\,\Sigma_{\bm{f}}(t';\bm{u}) \ .
\end{equation}
The LR contribution to the excess work is then
\begin{equation} \label{eq:LRexcesswork}
    \left\langle\mathcal{W}_{\text{ex}}\right\rangle^{\text{(LR)}} = \int_0^\tau\md t\frac{\md \bm{u}^{\mathsf{T}}}{\md t} \tilde{\zeta}(\bm{u}(t))\frac{\md \bm{u}}{\md t} \ .    
\end{equation}
The component $\tilde{\zeta}_{ij}(\bm{u})$ of the generalized friction tensor approximates how the energy cost of simultaneously changing the control parameters $u^i$ and $u^j$ scales with the velocity of the control parameters. While the exact excess work (\ref{eq:exactexcessworkV},\ref{eq:exactexcesswork}) depends on the solution $p(t)$ of the time-inhomogeneous equation (\ref{eq:MEnFPE}), the linear-response approximation~(\ref{eq:LRexcesswork}) only requires information about the equilibrium dynamics for fixed parameters. The excess work $\langle \mathcal{W}_{\text{ex}}\rangle$ approaches the linear-response approximation $\langle\mathcal{W}_{\text{ex}}\rangle^{(\text{LR})}$ 
as the protocol duration $\tau$ increases (Fig. ~\ref{fig:workConvergence}).

The generalized friction tensor is positive-definite and symmetric, endowing the control-parameter space with a Riemannian metric structure. Within this geometric framework, control protocols are curves on a \textit{thermodynamic manifold} $\tilde{\mathcal{M}}$ on which distances quantify the excess work for near-equilibrium driving. The LR excess work~(\ref{eq:LRexcesswork}) of a protocol $\bm{u}(t)$ is identified with the geometric energy of a curve $\bm{u}(t)$ on $\mathcal{M}$, and therefore LR minimum-work protocols are geodesics (paths of shortest length) on the thermodynamic manifold. Such protocols thus satisfy the geodesic equation $\nabla_{\dot{\bm{u}}}\dot{\bm{u}}=0$ for covariant derivative $\nabla_{\dot{\bm{u}}}$ along the curve. In components,
\begin{equation} \label{eq:geodesicequation}
    \frac{\md ^2u^i}{\md t^2} + \sum_{j, k}\Gamma^i_{\phantom{i}jk}\frac{\md u^j}{\md t}\frac{\md u^k}{\md t} = 0, \ \ \ i = 1, \dots, m \ ,
\end{equation}
for Christoffel symbols of the second kind 
\begin{equation} \label{eq:christoffeltwo}
    \Gamma^i_{\phantom{i}jk} \equiv \frac{1}{2}\sum_{\ell}\tilde{\zeta}^{i\ell}\left(\frac{\partial \tilde{\zeta}_{\ell j}}{\partial u^k} + \frac{\partial \tilde{\zeta}_{k \ell}}{\partial u^j} - \frac{\partial \tilde{\zeta}_{j k}}{\partial u^\ell} \right)    
\end{equation}
for $\tilde{\zeta}^{i\ell} \equiv \sbrac{\tilde{\zeta}^{-1}}_{i\ell}$~\cite{prasolovDifferentialGeometry2022}.  

In a small handful of systems~\cite{sivakThermodynamicMetricsOptimal2012, zulkowskiOptimalFinitetimeErasure2014, frimOptimalFinitetimeBrownian2022}, the friction tensor is analytically obtainable directly from the definition (\ref{eq:frictensdef}). In the special case of one-dimensional continuous overdamped dynamics, the friction tensor simplifies~\cite{zulkowskiOptimalControlOverdamped2015}, extending the range of analytically tractable models. Physically motivated approximations may also provide analytic insights when the full friction tensor is not known~\cite{deffnerKibbleZurekScalingIrreversible2017}. More often, numerical methods are required, typically involving estimation of the time-correlation functions $\Sigma_{\bm{f}}$ using Markov-chain Monte Carlo simulations~\cite{rotskoffOptimalControlNonequilibrium2015, louwerseMultidimensionalMinimumworkControl2022, guptaEfficientControlProtocols2023}.

\section{Hierarchy of Thermodynamic Manifolds}
\begin{figure}[t]
    \includegraphics[width = 3in]{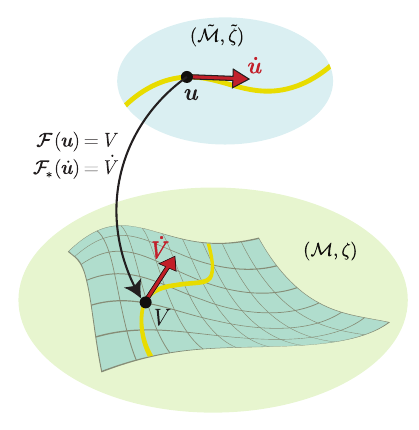}
    \caption{Partial-control manifolds $(\tilde{\mathcal{M}}, \tilde{\zeta})$ (blue) coordinatized by a set of control parameters $\bm{u}$ are submanifolds of the global thermodynamic manifold $(\mathcal{M},\zeta)$, with $\tilde{\zeta}$ inherited from $\zeta$ by Eq.~\eqref{eq:inheritedMetric}. The restriction $V = V(\bm{u})$ of conservative control defines the immersion $\mathcal{F}$~\eqref{eq:immersion}. The pushforward $\mathcal{F}_*$---that maps vectors on $\tilde{\mathcal{M}}$ to vectors on $\mathcal{M}$ (red arrows)---is related to the conjugate forces $\bm{f}$~\eqref{eq:conjforcesPushforwardDisc}.} 
    \label{fig:inducedmetric}
\end{figure}

In this section, we apply the mathematics of submanifolds~\footnote{While this section is intended to be self-contained, the essential geometric ideas will be unfamiliar to many readers. For a thorough overview of differential geometry including submanifolds and degenerate metrics with notation that will be familiar to readers versed in general relativity, see~\cite{eisenhart_geometry}. For a more modern treatment of the subject, see~\cite{CHEN2000187}.}
to connect the geometries of distinct control-parameter sets. By constructing a global thermodynamic manifold $\mathcal{M}$ spanned by the state energies, we show that the partial-control manifolds for arbitrary sets $\curly{u^0,u^1,\dots,u^{m-1}}$ of control parameters arise naturally as submanifolds~(Fig. \ref{fig:inducedmetric}). In conjunction with the decompositions we will derive in Sec.~\ref{sec:decomps}, this geometric picture offers methods for rapidly computing friction tensors.

\subsection{Global manifold: Full control \label{ssec:globalManifold}} 
In the context of conservative driving, `full control' refers to the ability to arbitrarily reshape the potential-energy landscape $V(x)$ provided that $\e{-\beta V(x)}$ remains normalizable. Equivalently, this means that the control-parameter set is sufficiently robust to prepare the controlled system in any equilibrium distribution $\pi$.

For a system with $N$ discrete states, the energy landscape is specified by the energy of all $N$ states, which can be assembled into a vector $\bm{V} = (V^0, V^1, \dots, V^{N-1})$ with $V^{\nu} = V(x_\nu)$. For a continuous system, one may imagine a continuum of control parameters $V(x)$ parameterized by $x$. The forces conjugate to the state energies are indicator functions
\begin{equation} \label{eq:fcconjforces}
     \omega(x,t) \equiv -\fdv{V(X_t)}{V(x)} = -\delta(x, X_t) \, ,
\end{equation}
where the functional derivative accounts for continuous $x$. The average deviation of the conjugate forces $\omega$ from their equilibrium values is then precisely the (negative) deviation of the probability distribution from equilibrium:
\begin{equation} \label{eq:occupaverage}
    \langle \delta \omega(x,t) \rangle \equiv \angbr{\delta \omega(x)}_{p(t)} = -\delta p(x, t) \ .
\end{equation}

Analogous to \eqref{eq:frictensdef}, we define the full-control friction tensor $\zeta(V)$---which depends on the entire potential-energy landscape $V$---by its integral kernel
\begin{equation} \label{eq:fullconttensordef}
    \zeta(x, y; V) \equiv  \beta \int_0^\infty \md t' \angbr{\delta \omega(x, t')\, \delta \omega(y, 0)}_{\text{eq}} \ , 
\end{equation}
such that
\begin{subequations} \label{eq:LRpower}
\begin{align}
    \angbr{\mathcal{P}_{\text{ex}}(t)} &= \int_{\Omega}\md \mu(x) \, \frac{\md V(x)}{\md t} \, \delta p(x,t) \\
    &\approx \int_{\Omega\times\Omega}\md\mu(x)\, \md\mu(y) \, \frac{\md V(x)}{\md t} \, \zeta(x,y;V) \, \frac{\md V(y)}{\md t} \\
    &\equiv \angbr{\mathcal{P}_{\text{ex}}(t)}^{(\text{LR})} \ .
\end{align}
\end{subequations} 
Interpreted geometrically, the full-control friction tensor is a metric on a (perhaps infinite-dimensional) manifold that we call the \textit{global thermodynamic manifold} $\mathcal{M}$, with each possible landscape identified with a point on $\mathcal{M}$. The squared length of a tangent vector on the global manifold corresponds to the linear-response excess power:
\begin{equation} \label{eq:velocityVector}
    \angbr{\mathcal{P}_{\text{ex}}(t)}^{(\text{LR})} =\paren{\frac{\md V}{\md t} , \zeta\frac{\md V}{\md t}} \equiv \left|\left|\frac{\md V}{\md t} \right|\right|_{\zeta}^2 \ ,
\end{equation}
where $||\cdot||_{\zeta}$ denotes a norm with respect to the metric $\zeta$. 

We will see later that the metric $\zeta$ has a non-empty null space spanned by $1$ (respectively the constant function $1$ for continuous $\Omega$ or the constant vector $\bm{1}$ for discrete $\Omega$). We should expect this intuitively from the linear-response relation implied by Eq.~(\ref{eq:LRpower}):
\begin{equation} \label{eq:LRrelation}
    \delta p \approx \zeta \frac{\md V}{\md t} \ .
\end{equation}
Uniformly raising or lowering the energy landscape does not affect the system's dynamics, so this action does not alter the system distribution: $\delta p = 0$ for $\md V/\md t \propto 1$. An equivalent argument may be made from conservation of probability \footnote{Taking the inner product of both sides of~\eqref{eq:LRrelation} with the constant function $1$ and applying $(1,\delta p) = 0$ leads to the same conclusion. Since $\zeta$ is symmetric, in this context the arbitrariness of the potential-energy zero-point and conservation of probability are equivalent. This highlights a duality between energy and probability, manifest in the geometry: the linear-response relation \eqref{eq:LRrelation} states that $\delta p$ is the vector dual to $\md V / \md t$. This is also reflected in the conjugate forces to the state energies~\eqref{eq:fcconjforces}.}.

Since $\text{det} \, \zeta = 0$, the metric is said to be \textit{degenerate} and the full-control manifold $\mathcal{M}$ is thus a non-Riemannian manifold. This degeneracy does not meaningfully impede the usual geometric analysis, however, as the null vector (in the coordinates of $V$) is the same everywhere on the manifold. The degeneracy may be managed by projecting points $V \in \mathcal{M}$ onto some hypersurface on which the metric is non-degenerate, e.g., by fixing the potential zero-point, setting $V(y) = 0$ for some $y \in \Omega$. We address the construction of Riemannian full-control submanifolds for discrete-state systems in Sec.~\ref{ssec:riemFCmanifold}. 

\subsection{Submanifolds: Partial control \label{ssec:submanifolds}}
In practical settings, the number of control parameters is significantly smaller than the number of system states. For instance, early studies of fluctuation theorems involved the manipulation of single RNA molecules with the molecular extension as the only controllable parameter~\cite{liphardtEquilibriumInformationNonequilibrium2002, collinVerificationCrooksFluctuation2005}. The success of these demonstrations and other nonequilibrium measurements of equilibrium free-energy landscapes relies on executing experimental protocols with low excess work~\cite{blaberOptimalControlStochastic2023}. Partial (or parametric) control is thus an area of clear practical importance.

For conservative control with a set of $m$ control parameters $\curly{u^0, \dots, u^i,\dots,u^{m-1}}$, protocols are constrained by $V(t)=V(\bm u(t))$ to a lower-dimensional control-parameter space. Geometrically, this space is a submanifold $\tilde{\mathcal{M}}$ of the global manifold $\mathcal{M}$. 

Since the control parameters represent a constrained control over the potential energy $V$, there is a map (unique up to specification of the energy zero point) between points $\bm{u}$ on $\tilde{\mathcal{M}}$ and points $V(\bm{u})$ on $\mathcal{M}$. In differential geometry, such a map is an \textit{immersion}: 
\begin{align} \label{eq:immersion}
    \begin{split}
        \mathcal{F} &: \tilde{\mathcal{M}} \to \mathcal{M} \\
        &: \boldsymbol{u} \mapsto V(\boldsymbol{u})
    \end{split}
\end{align}
The partial-control manifold is called a submanifold of the global manifold with immersion $\mathcal{F}$. 

Associated with an immersion is its \textit{pushforward} (or \textit{differential}) $\mathcal{F}_* : \text{T}_{\bm u}\tilde{\mathcal{M}}\to \text{T}_{V(\bm{u})}\mathcal{M}$, mapping vectors on $\tilde{\mathcal{M}}$ (i.e., elements of the tangent space $\text{T}_{\bm u}\tilde{\mathcal{M}}$) to vectors on $\mathcal{M}$. In this setting, the matrix elements of the pushforward are simply obtained by the chain rule, and are identified with the forces $f_i(x)$ conjugate to respective parameters $u^i$ when the system is in state $x$:
\begin{equation} \label{eq:conjforcesPushforwardDisc}
    \dot{V}(x) =\sbrac{\mathcal{F}_*(\dot{\bm{u}})}(x) = -\sum_{i=0}^{m-1}f_{i}(x) \,\dot{u}^i \ .
\end{equation}

The metric $\tilde{\zeta}(\bm{u})$ on $\tilde{\mathcal{M}}$, inherited from $\zeta(V)$, is then 
\begin{equation} \label{eq:inheritedMetric}
    \tilde{\zeta}_{ij}(\boldsymbol{u}) = \int_{\Omega \times \Omega}\md\mu(x)\, \md\mu(y) \, \zeta(x,y;V(\bm{u})) \, f_i(x)f_{j}(y) \ .
\end{equation}
Substituting \eqref{eq:fullconttensordef} into this form and applying \eqref{eq:fcconjforces} shows it to indeed be equivalent to the metric defined in Eq.~(\ref{eq:frictensdef}). This is constructed to ensure that geometric invariants (e.g., the lengths of tangent vectors) are equal when computed on $\mathcal{M}$ and on $\tilde{\mathcal{M}}$. In particular, since the squared length of a vector on a thermodynamic manifold corresponds to the linear-response excess power, the preservation of geometric invariants under coordinate transformations means that one computes the same excess power from $\tilde{\zeta}$ and $\bm{u}(t)$ as from $\zeta$ and $V(t)$. Figure~\ref{fig:inducedmetric} summarizes this picture.

\subsection{Riemannian full-control submanifolds \label{ssec:riemFCmanifold}}
For a discrete-state system with $|\Omega| = n$, full control is possible with a set of $n - 1$ parameters due to the invariance of the dynamics under a uniform shift of the state energies (captured in the degeneracy of $\zeta$). We define a control-parameter set $\bm{u}\in\mathbb{R}^{n-1}$ as \textit{complete} if the potential energy is fully determined (up to a global additive constant) by $\bm{u}$. 

This may be formally stated as follows: a control-parameter set $\bm{u}$ is complete if for all $\bm{V}' \in \mathbb{R}^n$, there exists some $\bm{u}'\in\mathbb{R}^{n-1}$ and $c \in \mathbb{R}$ such that
\begin{equation} \label{eq:completenessFormal} 
\bm{V}' = \bm{V}(\bm{u}') + c\bm{1} \ .
\end{equation}
This implies a sort of point-wise generalized invertibility between $\bm{u}$ and $\bm{V}$. If, in addition, $\bm{u}'$ is \textit{unique} for each $\bm{V}'$, then we may always project the hypersurface defined by $\bm{V}(\bm{u})$ onto a hyperplane, associating to each control vector a unique energy vector in the plane. 

Here we consider complete control sets $\bm{u}$ that only affect relative energies and not the overall offset, so $\bm{V}(\bm{u})$ is a hyperplane with constant normal vector $\bm{1}$.

Define the Jacobian $\mathbb{J} = \mathcal{F}_*^{\mathsf{T}}$ of the transformation from the control parameters $\bm{u}$ to the state energies $\bm{V}$, i.e., $\sbrac{\mathbb{J}}^{\phantom{i}\nu}_{i} = -f_i(x_{\nu}) = \partial V^{\nu}(\bm{u})/\partial u^i$. Then for probability vector $\bm{p}(t)$,
\begin{equation} \label{eq:avgFJac}
    \angbr{\bm{f}}_{p(t)} = -\mathbb{J}\bm{p}(t) \ . 
\end{equation}
We show in App.~\ref{app:fcc} that under the stated conditions, the distribution $\bm{p}(t)$ is completely determined by the average conjugate forces $\angbr{\bm{f}}_{\bm{p}(t)}$, with Eq.~\eqref{eq:avgFJac} invertible via
\begin{equation} \label{eq:completenessCondition2}
    \bm{p}(t) = - \mathbb{J}^{\mathsf{R}}\angbr{\bm{f}(t)}+\frac{1}{n}\bm{1} 
\end{equation}
for right inverse $\mathbb{J}^{\mathsf{R}} = \mathbb{J}^{\mathsf{T}}\paren{\mathbb{J}\mathbb{J}^{\mathsf{T}}}^{-1}$.

The time evolution of the average deviation $\angbr{\delta \bm{f}(t)}$ of the conjugate force to a complete set of control parameters is thus self-contained in the sense that 
\begin{equation}
    \frac{\md \angbr{\delta \bm{f}(t)}}{\md t} = \mathbb{W}_{\bm{f}}\angbr{\delta \bm{f}(t)}
\end{equation}
where 
\begin{equation} \label{eq:generatorF}
    \mathbb{W}_{\bm{f}} \equiv \mathbb{J}\mathbb{W}\mathbb{J}^{\mathsf{R}} 
\end{equation}
is the generator of the dynamics of $\angbr{\bm{f}(t)}$. 

\begin{figure} 
    \centering 
    \includegraphics[width=\linewidth]{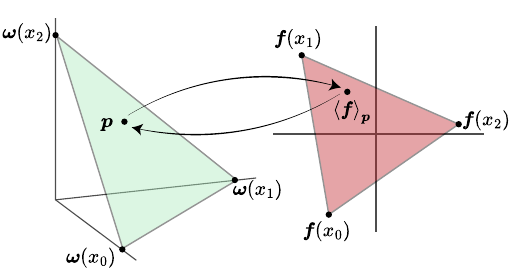}
    \caption{Schematic of the geometric condition on the completeness of a set of control parameters $\bm{u}$. A control parameter set $\bm{u}$ is \textit{complete}, i.e., able to fully determine state energies up to a constant offset when the space $\Delta_f$ of average conjugate forces $\angbr{\bm{f}}$ forms an $(n-1)$-simplex in $\mathbb{R}^{n-1}$ (red triangle, right), since this allows a bijective transformation between $\Delta_{\bm{f}}$ and the probability simplex $\Delta_{\bm{p}}$ (green triangle, left).} 
    \label{fig:affinegeometry}
\end{figure}

There is an equivalent geometric condition on the completeness of a control-parameter set, illustrated for a three-state system in Fig.~\ref{fig:affinegeometry}. Let $\Delta_{\bm{f}}$ be the convex hull of the $n$ points $\curly{\bm{f}(x_i)}$, i.e., the set of all conjugate-force averages $\angbr{\bm{f}}$. For a complete set of control parameters, $\Delta_f$ forms an $(n-1)$-dimensional simplex in $\mathbb{R}^{n-1}$. Since the space of probability vectors $\bm{p}$ for an $n$-state system defines an $(n-1)$-dimensional simplex (the \textit{probability simplex} $\Delta_{\bm{p}} = \curly{\bm{p}\in\mathbb{R}^n \ : \ \bm{1}^{\mathsf{T}}\bm{p} = 1}$), there exists a bijective transformation between these spaces. Put simply, the space of conjugate-force averages $\angbr{\bm{f}}$ for a complete set of control parameters is the same ``size'' as the space of probability vectors. 

Concretely, consider an $N$-spin system, for which there are $n = 2^N$ states. We show in Sec.~\ref{ssec:spinSystems} that the set of all $k$-spin interactions for $k = 1,\dots,N$ from the cluster expansion is a complete control set that spans a $(2^N - 1)$-dimensional Riemannian full-control submanifold.

\section{Decompositions of the friction tensor \label{sec:decomps}} 
This hierarchy of thermodynamic manifolds means that, given full-control metric $\zeta$, we may compute a partial-control metric $\tilde{\zeta}$ for \textit{any} control-parameter set $\curly{u^0,\dots,u^{m-1}}$. Though conceptually appealing, this leaves open the question of how we can actually calculate and physically interpret the friction tensor $\zeta$ and its relationship to inherited metrics $\tilde{\zeta}$. In this section, we show that the properties of the full-control friction tensor (\ref{eq:fullconttensordef}) permit mathematical manipulations which significantly simplify the calculation of friction tensors, offer a dynamical interpretation of the linear-response excess power, and suggest design principles for reducing dissipation in driven systems. 

\subsection{The Drazin inverse \label{subsec:group}} 
We begin with a brief review of the Drazin inverse, a generalized inverse for linear operators which plays a central role in the derivations that follow. The Drazin inverse of the rate matrix has also appeared in previous works on optimal control~\cite{crooksDrazinInverseRate2018, mandalAnalysisSlowTransitions2016, scandiThermodynamicLengthOpen2019,millerWorkFluctuationsSlow2019, zhongLinearResponseEquivalence2024} and has applications in first-passage problems~\cite{yaoFirstpassagetimeMomentsMarkov1985, coolen-schrijnerDeviationMatrixContinuoustime2002}.  

Intuitively, the need for an appropriately defined inverse of the generator arises due to the time integral in the definition~\eqref{eq:completeControlFriction} of the full-control friction tensor. In the same way as the generator $\mathcal{G}_t$ plays the role of a time derivative on the space of probability distributions, its Drazin inverse $\mathcal{G}^{\mathcal{D}}_t$ plays the role of a time integral (over $[0,\infty)$) on the same space. 

Let $\mathcal{A}$ be a linear operator on some Hilbert space. The Drazin inverse of $\mathcal{A}$ is the unique operator $\mathcal{A}^{\mathcal{D}}$ satisfying
\begin{subequations} \label{eq:drazinprops}
    \begin{align} \label{eq:drazinprop1}
     \mathcal{A}\mathcal{A}^{\mathcal{D}}\mathcal{A}^{k} &= \mathcal{A}^{k} \\ 
     \mathcal{A}^{\mathcal{D}}\mathcal{A}\mathcal{A}^{\mathcal{D}} &= \mathcal{A}^{\mathcal{D}}  \label{eq:drazinprop2} \\ 
     \mathcal{A}^{\mathcal{D}}\mathcal{A} &= \mathcal{A}\mathcal{A}^{\mathcal{D}} \ , \label{eq:drazinprop3}
\end{align}
\end{subequations}
for some $k \in \mathbb{Z}$ called the index of $\mathcal{A}$. For operators on Hilbert spaces of finite dimension (i.e., matrices), $k$ is the smallest non-negative integer such that $\text{rank} \, \mathcal{A}^k = \text{rank} \, \mathcal{A}^{\ell}$ for all $\ell > k$~\cite{wangGeneralizedInversesTheory2018}. We consider only the case of $k = 1$ (such as for irreducible rate matrices~\cite{vankampenMasterEquation2007}). In this case, the Drazin inverse of $\mathcal{A}$ may be constructed from
\begin{equation} \label{eq:drazininverse}
    \mathcal{A}^{\mathcal{D}} = P_0 + (\mathcal{A} - P_0)^{-1} \ ,
\end{equation}
for projection matrix $P_0$ onto the null space $\mathcal{N}(\mathcal{A})$ of $\mathcal{A}$. 

From properties (\ref{eq:drazinprops}a-c), one can show that
\begin{equation} \label{eq:drazinProjection}
    \mathcal{A}\mathcal{A}^{\mathcal{D}} = I - P_0 \ ,
\end{equation}
for identity operator $I$. That is, $\mathcal{A}\mathcal{A}^{\mathcal{D}}$ projects vectors in the Hilbert space onto the complement of $\mathcal{N}(\mathcal{A})$.

Concretely, suppose that $\mathbb{W}$ is an irreducible rate matrix with equilibrium distribution $\bm{\pi}$. Then the product $\mathbb{W}\mathbb{W}^{\mathcal{D}}$ (and $\mathbb{W}^{\mathcal{D}}\mathbb{W}$ by the commutativity property~\eqref{eq:drazinprop3}) maps a probability distribution to its deviation from equilibrium, i.e., $\mathbb{W}\mathbb{W}^{\mathcal{D}}\bm{p} = (I - \Pi)\bm{p} = \delta\bm{p}$ with $\Pi = \bm{\pi}\bm{1}^{\mathsf{T}} = \sbrac{\bm{\pi}\,\cdots\,\bm{\pi}}$. As a corollary, 
\begin{equation} \label{eq:DrazinDevInv}
    \mathbb{W}\mathbb{W}^{\mathcal{D}}\delta \bm{p} = \delta\bm{p} \ ,
\end{equation} 
so that $\mathbb{W}^{\mathcal{D}}$ is a proper inverse of $\mathbb{W}$ when the domain is restricted to vectors in the hyperplane $\curly{\bm{v} \ : \ \bm{1}^{T}\bm{v} = 0}$ -- including deviations $\delta\bm{p}$ from equilibrium and derivatives $\md \bm{p}/\md t$ of probability distributions.

From properties~(\ref{eq:drazinprops}b,c), it follows that \footnote{Property~\eqref{eq:drazinprop2} can be re-written as $(\mathcal{A}^{\mathcal{D}})^2\mathcal{A} = \mathcal{A}^{\mathcal{D}}$ by the commutative property~\eqref{eq:drazinprop3}. Thus $\mathcal{A}x = 0 \implies \mathcal{A}^{\mathcal{D}}x = 0$. An identical argument holds for right-multiplication.}
\begin{equation} \label{eq:drazinnullspace}
    \mathbb{W}^{\mathcal{D}}\bm{\pi} = \bm{1}^{\mathsf{T}}\mathbb{W}^{\mathcal{D}} = 0 \ .
\end{equation}
The Drazin inverse may thus be interpreted as ``inverting the invertible part'' of a matrix~\cite{crooksDrazinInverseRate2018}, leaving the null space of $\mathbb{W}^{\mathcal{D}}$ coincident with the null space of $\mathbb{W}$. 

\subsection{Operator decomposition} 
Consider some continuous- or discrete-state Markov process with generator $\mathcal{G}$ satisfying the conditions established in Sec.~\ref{sec:prelims}. We would like to manipulate the expression~(\ref{eq:fullconttensordef}) for the full-control friction tensor into a more manageable form. We take advantage of the fact that the integrand $\Sigma_{\omega}(t) = \angbr{\delta \omega(t) \delta\omega(0)}_{\text{eq}}$ is calculated for a fixed energy landscape so that $\md \pi / \md t = 0$ and thus $\md \Sigma_{\omega}/\md t = \mathcal{G}\Sigma_{\omega}$. Then from~(\ref{eq:DrazinDevInv}), the full-control friction tensor is 
\begin{subequations}
    \begin{align}
        \zeta &= \beta \int_0^{\infty}\md t' \, \Sigma_{\omega}(t') \\
        &= \beta \int_0^{\infty}\md t' \, \mathcal{G}^{\mathcal{D}}\frac{\md \Sigma_{\omega}(t')}{\md t} \\
        &= -\beta\, \mathcal{G}^{\mathcal{D}}\,\Sigma_{\omega}^0 \label{eq:dyndecomp1}
    \end{align}
\end{subequations}
for $\Sigma_{\omega}^0 \equiv \Sigma_{\omega}(0)$. The final line follows from the fact that $\lim_{t\to\infty} \Sigma_{\omega}(t) = 0$ for ergodic dynamics. 

Equation~(\ref{eq:dyndecomp1}) is a main result: the full-control friction tensor is decomposable into a product of the Drazin inverse of the generator and the equal-time equilibrium covariance matrix of the indicator functions $\omega$. Roughly, this separates the friction tensor into a static part $\Sigma_{\omega}^0$ and a dynamic part $\beta\mathcal{G}^{\mathcal{D}}$. 

This expression simplifies further. Appendix~\ref{app:eqcov} shows that the equal-time equilibrium covariance matrix of the indicator functions $\omega$ satisfies
\begin{equation} 
    \Sigma_{\omega}^0(x,y) = \pi(x)\delta(x,y) - \pi(x)\pi(y) \ .
\end{equation}
Since $\pi$ is in the null-space of $\mathcal{G}^{\mathcal{D}}$, we obtain the decomposition
\begin{equation} \label{eq:dynamicalDecomposition}
    \zeta(x,y) = -\beta\,\mathcal{G}^{\mathcal{D}}(x,y) \,\pi(y) \ . 
\end{equation}
For discrete systems, this reduces the infinite-time integral (\ref{eq:fullconttensordef}) to a matrix equation
\begin{equation} \label{eq:dynamicalDecompositionDiscrete}
    \zeta = -\beta \mathbb{W}^{\mathcal{D}}\text{diag}\curly{\bm{\pi}} \ .
\end{equation}
Together with the formula~(\ref{eq:drazininverse}) for the Drazin inverse of a matrix, this expression allows for machine-precision numerical calculation of the full-control friction tensor $\zeta(V)$ -- and thus \textit{any} partial-control tensor $\tilde{\zeta}(\bm{u})$ via the analysis of Sec.~\ref{ssec:submanifolds}.

In App.~\ref{app:fullControlFriction}, we show that for a complete control set $\bm{u}$, the friction tensor is
\begin{equation} \label{eq:completeControlFriction}
    \tilde{\zeta}(\bm{u}) = -\beta \mathbb{W}^{-1}_{\bm{f}}\Sigma_{\bm{f}}^0 
\end{equation}
for generator $\mathbb{W}_{\bm{f}}$ of the dynamics of $\angbr{\bm{f}(t)}$ as defined in Eq.~(\ref{eq:generatorF}) and equal-time correlation matrix $\Sigma_{\bm{f}}^0 = \angbr{\delta\bm{f}\delta\bm{f}^{\mathsf{T}}}_{\bm{\pi}(\bm{u})}$. 

Finally, since a protocol $V(t)$ can be completely characterized up to an (irrelevant) additive constant by the evolution of the equilibrium distribution $\pi(t)$, we use the operator decomposition to derive the relation (App.~\ref{app:fricDrazProof}) 
\begin{equation}\label{eq:fricDrazRelation}
    \zeta \frac{\md V}{\md t} = \mathcal{G}^{\mathcal{D}}\frac{\md \pi}{\md t} \ .
\end{equation}
The linear-response relation~\eqref{eq:LRrelation} is then equivalent to the approximation
\begin{equation}
    \delta p \approx \mathcal{G}^{\mathcal{D}}\frac{\md \pi}{\md t} \ .
\end{equation}
This expression was derived previously using an operator expansion~\cite{mandalAnalysisSlowTransitions2016}.

\subsection{Connection to integral relaxation time}
As noted in Ref.~\cite{sivakThermodynamicMetricsOptimal2012}, the friction tensor $\tilde{\zeta}$ (for any set of control parameters) may be written as 
\begin{equation} \label{eq:altdecomp} 
    \tilde{\zeta} = k_{\rm B} T \ \mathcal{T}\circ \mathcal{I} \ ,
\end{equation}
the Hadamard product of the integral relaxation time matrix 
\begin{equation}
    \mathcal{T}_{ij} \equiv \int_0^\infty \md{t'} \frac{\angbr{\delta f_i(t')\delta f_j(0)}_{\text{eq}}}{\angbr{\delta f_i \delta f_j}_{\bm{\pi}}}
\end{equation}
and the equilibrium Fisher-information matrix
\begin{equation}
    \mathcal{I} \equiv \beta^2 \Sigma_{\bm{f}}^0 \ .
\end{equation}
When the conjugate forces $\bm{f}$ are the indicator functions $\bm{\omega}$, equating (\ref{eq:altdecomp}) with the operator decomposition (\ref{eq:dynamicalDecompositionDiscrete}) gives a simple correspondence between the Drazin inverse of the rate matrix and the integral relaxation time:
\begin{equation}
    \mathbb{W}_{\mu\nu}^{\mathcal{D}} = \begin{dcases} \pi_\mu\mathcal{T}_{\mu\nu} \ , \ \ &  \mu\neq\nu  \\
    -(1 - \pi_\mu)\mathcal{T}_{\mu\mu} \ , & \mu = \nu \ .
    \end{dcases}
\end{equation}
Seen another way, defining the transition probability matrix with elements $p_{\mu\nu}(t) = \text{Pr}\sbrac{\mu, t \ | \ \nu, 0}$, the Drazin inverse of the rate matrix is~\cite{coolen-schrijnerDeviationMatrixContinuoustime2002}
\begin{subequations}
\begin{align} \label{eq:altdrazin} 
    -\mathbb{W}_{\mu\nu}^{\mathcal{D}} &= \int_0^{\infty}d t \, \sbrac{p_{\mu\nu}(t) - \pi_\mu} \ .
\end{align}
\end{subequations}
Thus, much like $\mathcal{T}$, the Drazin inverse of the rate matrix describes the timescale of system relaxation. The magnitude of $\mathbb{W}^{\mathcal{D}}_{\mu\nu}$ characterizes how long it takes for $p_{\mu\nu}(t)$, the probability $p_\mu(t)$ conditioned on initialization in state $\nu$, to reach the equilibrium probability $\pi_\mu$.

\subsection{Spectral decomposition \label{ssec:spectralDecomp}}
When a generator $\mathcal{G}$ admits the spectral decomposition~(\ref{eq:generatorSpectralDecomposition}), the integral kernel of the Drazin inverse is
\begin{equation}
    \mathcal{G}^{\mathcal{D}}(x,y) = -\sum_{\alpha\geq  1}\tau_{\alpha}\psi_{\alpha}(x)\phi_{\alpha}(y) 
\end{equation}
for relaxation times
\begin{equation}
    \tau_{\alpha} \equiv -\frac{1}{\lambda_{\alpha}} \ , \ \ \alpha\geq 1 \ .
\end{equation}
See App.~\ref{app:specDecDraz} for a concise proof. Substituting this expression into the decomposition (\ref{eq:dynamicalDecomposition}) and using the fact that $\psi_{\alpha}(x) = \phi_{\alpha}(x)\pi(x)$ for systems obeying detailed balance~\cite{riskenFokkerPlanckEquationMethods1996}, we obtain a spectral decomposition of the full-control friction tensor:
\begin{equation} \label{eq:spectralDecomposition}
    \zeta(x,y) = \beta\sum_{\alpha \geq 1} \tau_{\alpha}\psi_{\alpha}(x)\psi_{\alpha}(y) \ .
\end{equation}
The degeneracy of $\zeta$ as inferred from Eq.~\eqref{eq:LRrelation} is manifest in this form: from the biorthogonality of the eigenfunctions, we have $\phi_0 \zeta = \zeta \phi_0 = 0$, with $\phi_0 = 1$.

This spectral decomposition can be used to express the linear-response excess power~(\ref{eq:LRpower}) as a sum of independent contributions associated with the fixed-parameter relaxation modes of the system:
\begin{equation} \label{eq:spectralPower}
    \langle\mathcal{P}_{\text{ex}}(t)\rangle^{(\text{LR)}} = \beta\sum_{\alpha \geq 1} \tau_{\alpha}(t)\paren{\frac{\md V(t)}{\md t}, \, \psi_{\alpha}(t)}^2 \ , 
\end{equation}
Under the change of variables $s \equiv t/\tau$ for total protocol time $\tau$, the linear-response excess work is 
\begin{equation} \label{eq:spectralLRWork}
        \angbr{\mathcal{W}_{\text{ex}}}^{(\text{LR})}(\tau) = \frac{k_{\rm B} T}{\tau} \sum_{\alpha \geq 1} \int_0^1 \md s \, \tau_{\alpha} \paren{\frac{\md \beta V}{\md s}, \,\psi_\alpha}^2  \ .
\end{equation} 
Equations~\eqref{eq:spectralDecomposition}-\eqref{eq:spectralLRWork} constitute another main result. 
The spectral decomposition of the excess work provides insight into the dynamical origin of dissipation: roughly, each term captures the degree to which the protocol traverses a given mode, scaled by that mode's relaxation time. As a design principle, dissipation may be reduced by driving a system such that $\md V/\md t$ is maximally orthogonal to modes with long relaxation times.

A similar expression holds for partial-control submanifolds. From the transformation law for metrics,
\begin{equation} \label{eq:frictionSpectralPartial}
    \tilde{\zeta}_{ij}(\bm{u}) = \sum_{\alpha \geq 1} \tau_{\alpha} \, \tilde{\psi}_{\alpha|i} \, \tilde{\psi}_{\alpha|j}
\end{equation}
for conjugate relaxation modes
\begin{equation} \label{eq:conjModes}
    \tilde{\psi}_{\alpha|i} = -\int_{\Omega}\md\mu(x)\,\psi_{\alpha}(x)f_i(x) \ .
\end{equation}
The zeroth conjugate relaxation mode $\tilde{\bm{\psi}}_0$ is the equilibrium average $-\angbr{\bm{f}}_{\pi}$ of the conjugate-force vector, and for $\alpha \geq 1$, $\tilde{\bm{\psi}}_\alpha$ captures the fixed-parameter relaxation of $-\angbr{\delta \bm{f}(t)}$ corresponding to system relaxation along mode $\alpha$. The linear-response excess power in terms of a limited set of control parameters is then
\begin{equation} \label{eq:powerPartialControl}
    \angbr{\mathcal{P}_{\text{ex}}(t)}^{(\text{LR})} = \beta\sum_{\alpha \geq 1}\tau_{\alpha}(t)\paren{\frac{\md \bm{u}(t)}{\md t}, \ \tilde{\bm{\psi}}_{\alpha}(t)}^2 \ .
\end{equation}
Equation~(\ref{eq:frictionSpectralPartial}) is particularly powerful for continuous systems for which the spectrum of $\mathcal{L}$ is analytically known. While the full-control spectral decomposition~(\ref{eq:spectralDecomposition}) is a manifestly infinite sum for continuous state spaces, the partial-control spectral decomposition~(\ref{eq:frictionSpectralPartial}) may be a finite sum if the conjugate forces $f_i(x)$ are orthogonal to all but a finite set of modes $\psi_{\alpha}(x)$; Section~(\ref{ssec:harmtrap}) demonstrates a concrete example of this.

\section{Theoretical synopsis \label{sec:preexSummary}}
We have, in the preceding sections, presented a rich geometric structure of dissipation under slow driving. Here we briefly review the essential features of this structure before applying it in three illustrative examples. 

To each dynamical system evolving reversibly and ergodically under the dynamics~\eqref{eq:MEnFPE}, one may define a \textit{global thermodynamic manifold} whose intrinsic geometry captures the relaxation dynamics, as made evident by the decomposition of the metric tensor in terms of the spectrum of the infinitesimal generator~\eqref{eq:spectralDecomposition}. This geometry also captures dissipation for slow driving: the squared length of the tangent vector approximates the excess power dissipation for a particular speed and direction of driving~\eqref{eq:velocityVector}. This generalizes the friction tensor defined in \cite{sivakThermodynamicMetricsOptimal2012}, and subsumes parameter-dependent friction tensors as inherited metrics~\eqref{eq:inheritedMetric} on submanifolds. The operator decomposition~\eqref{eq:dynamicalDecomposition} of the full-control metric identifies an object that has implicitly arisen in different forms in previous works \cite{mandalAnalysisSlowTransitions2016, crooksDrazinInverseRate2018, zhongLinearResponseEquivalence2024}, unifying its discrete- and continuous-state presentations.

We next demonstrate in examples the computational advantages conferred by operator decomposition and the submanifold structure, provide a concrete example of a Riemannian full-control manifold, and explore the analytical tools offered by the full- and partial-control spectral decompositions.

\section{Illustrative examples \label{sec:examples}}
\begin{figure}
    \centering
    \includegraphics[width=246 pt]{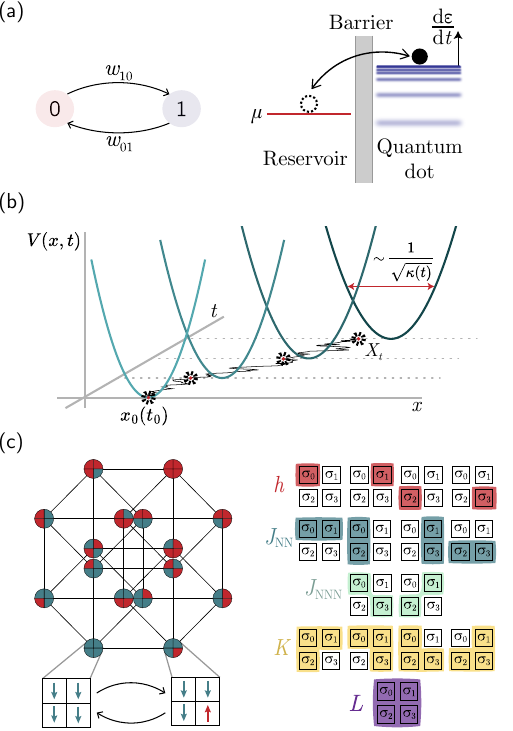}
    \caption{Model systems studied in Sec.~\ref{sec:examples}: (a) two-state system with transition rates $w_{01}$ and $w_{10}$ (left), one instantiation of which is a quantum dot interacting with a metallic reservoir~\cite{espositoFiniteTimeThermodynamics2010} (right); 
    (b) overdamped diffusion $X_t$ of a particle in a harmonic trap $V(x,t)$ parameterized by tunable stiffness $\kappa(t)$ and center $x_0(t)$; 
    (c) four-spin system with single spin-flip dynamics, giving rise to a state graph that is isomorphic to a hypercube (left). Control-parameter sets are 
    (possibly constrained)
    collections of $k$-spin couplings ($k = 1,2,3,4$) drawn from 15 parameters of the cluster expansion~\eqref{eq:clusterEnergy}, e.g., the 5-parameter set $(h, J_{\text{NN}}, J_{\text{NNN}}, K, L)$ where $h = h_0=h_1=h_2=h_3$, $J_{\text{NN}}=J_{01}=J_{02}=J_{13}=J_{23}$, $J_{\text{NNN}}=J_{03}=J_{12}$, and $K=K_{012}=K_{013}=K_{023}=K_{123}$
    (right).}
    \label{fig:modelSystems}
\end{figure}

\subsection{Two-state system \label{ssec:twoState}}

We begin with a very simple model system: two states with arbitrary transition rates $w_{10}$ and $w_{01}$ [\hyperref[fig:modelSystems]{Fig. 4(a)}, left]. From Eq.~(\ref{eq:drazininverse}), the Drazin inverse of the rate matrix is 
\begin{equation}
    \mathbb{W}^{\mathcal{D}} = \frac{1}{(w_{01} + w_{10})^2}\begin{bmatrix} 
        -w_{10} & w_{01} \\
        w_{10} & - w_{01}
    \end{bmatrix} \ .
\end{equation}
From the decomposition (\ref{eq:dynamicalDecompositionDiscrete}) and instantaneous equilibrium distribution $\paren{\pi_0(t), \pi_1(t)}^{\mathsf{T}} = \paren{\frac{w_{01}}{w_{01}+w_{10}}, \frac{w_{10}}{w_{01}+w_{10}}}^{\mathsf{T}}$ the full-control friction tensor is
\begin{equation} \label{eq:twostatefriction}
    \zeta = \beta\, \pi_0(t) \, \pi_1(t) \, \tau_{\text{rel}}(t) \begin{bmatrix}
        1 & -1 \\ -1 & 1
    \end{bmatrix} \ ,
\end{equation}
for relaxation time $\tau_{\text{rel}}(t) = (w_{01}(t)+w_{10}(t))^{-1}$. Note that the two-state system has only one non-zero eigenvalue (and thus only one relaxation time $\tau_{\text{rel}} \equiv \tau_1$). It may be readily verified that the spectral decomposition $\zeta(\bm{V}) = \beta \tau_{\text{rel}}\bm{\psi}_1\bm{\psi}_1^{\mathsf{T}}$ yields the same result.

The control parameters $\bm{V}$ may be freely defined subject to
\begin{equation} 
    -\beta(V^1 - V^0) = \ln\frac{\pi_1}{\pi_0} \ .
\end{equation}
Thus the energy difference $\Delta \equiv V^1 - V^0$ is a natural parameter to consider. The linear-response excess power is
\begin{equation} \label{eq:twostatePower}
    \angbr{\mathcal{P}_{\text{ex}}(t)}^{(\text{LR})} = \beta\pi_0(t)\pi_1(t)\tau_{\text{rel}}(t)\paren{\frac{\md \Delta}{\md t}}^2 
\end{equation}
and any LR minimum-work protocol $\Delta(t)$ satisfies 
\begin{equation}
    \frac{\md^2 \Delta}{\md t^2} + \frac{1}{2}\paren{\frac{\md \Delta}{\md t}}^2\frac{\partial\paren{\pi_0\pi_1\tau_{\text{rel}}}}{\partial \Delta} = 0  \ .
\end{equation}

A physical instantiation of a two-state system is a single-level quantum dot interacting with a metallic reservoir [\hyperref[fig:modelSystems]{Fig. 4(a)}, right]. In the wide-band approximation (i.e., the metallic reservoir's density of states is assumed constant~\cite{covitoTransientChargeEnergy2018}), the energy $\varepsilon$ of the quantum dot and the chemical potential $\mu$ of the reservoir set the respective tunneling rates $w_{10} = 2R_0[1+\e{\beta (\mu - \varepsilon)}]^{-1}$ into the reservoir and $w_{01} = 2R_0[1 + \e{\beta(\varepsilon - \mu)}]^{-1}$ out of the reservoir, where $R_0$ is the tunneling rate when $\mu = \varepsilon$~\cite{espositoFiniteTimeThermodynamics2010}. Then the friction tensor (for one control parameter, the friction \textit{coefficient}) is
\begin{equation}
    \tilde{\zeta}(\Delta) = \frac{\beta}{4 R_0}\sech^2 \paren{\tfrac{1}{2}\beta\Delta} \ ,
\end{equation}
for $\Delta = \varepsilon - \mu$. 

\subsection{Harmonic trap \label{ssec:harmtrap}}
Consider the overdamped diffusion $X_t$ of a particle in a one-dimensional harmonic trap $V(x,t) = \frac{1}{2}\kappa(t)(x - x_0(t))^2$ with stiffness $\kappa(t)$ and trap center $x_0(t)$ [\hyperref[fig:modelSystems]{Fig. 4(b)}]. The Fokker-Planck operator governing the evolution of the particle's position distribution $p(x,t)$ is
\begin{equation} 
    \mathcal{L}_t \, p(x,t) = \frac{1}{\beta \gamma}\frac{\partial}{\partial x}\curly{ \e{-\beta V(x, t)} \frac{\partial}{\partial x}\sbrac{\e{\beta V(x, t)} \ p(x,t)\ } } \ ,
\end{equation} 
for (Cartesian) friction coefficient $\gamma$ and inverse temperature $\beta$. Dropping the explicit time-dependence, the right eigenfunctions of $\mathcal{L}$ are~\cite{riskenFokkerPlanckEquationMethods1996} (after normalizing to $(\psi_{\alpha},\phi_{\alpha})=1$ with $\phi_{\alpha} =\psi_{\alpha}/\pi$)
\begin{equation} \label{eq:harmtrapeigfunc}
    \psi_{\alpha}(x) = (-1)^\alpha \sqrt{\frac{(\beta\kappa)^{1-\alpha}}{2\pi \alpha !}}\frac{\partial^\alpha}{\partial x^\alpha}\e{-\beta V(x)} \ ,
\end{equation} 
with relaxation times
\begin{equation} \label{eq:harmtrapeigval}
    \tau_\alpha = \frac{\gamma}{\alpha \kappa} \ .
\end{equation}

This is sufficient to write down an infinite sum for the full-control friction tensor $\zeta$ using~\eqref{eq:spectralDecomposition}. A more practical expression may be obtained for the metric $\tilde{\zeta}$ on the $(\kappa, x_0)$-thermodynamic submanifold: the conjugate forces $f_{x_0} = -\partial V / \partial x_0$ and $f_{\kappa} = -\partial V / \partial \kappa$ are determined by the first three left eigenfunctions:
\begin{subequations}
\begin{align}
    f_{x_0}(x) &= -\sqrt{\frac{\kappa}{\beta}} \, \phi_1(x) \ , \label{eq:cf1} \\
    f_{\kappa}(x) &= -\frac{1}{2\beta \kappa} \left[\phi_0(x) + \sqrt{2}\,\phi_2(x)\right] \label{eq:cf2}  \ .
\end{align}
\end{subequations}
Then by the orthonormality of the eigenfunctions, 
$(\phi_{\alpha},\psi_{\beta}) = \delta_{\alpha\beta}$,
the conjugate relaxation modes $\tilde{\psi}_{\alpha | i} = \paren{-f_i, \psi_{\alpha}}$~\eqref{eq:conjModes} are
\begin{subequations}
\begin{align}
    \tilde{\bm{\psi}}_1 &= \sqrt{\frac{\kappa}{\beta}}\begin{pmatrix}
        1 \\ 0
    \end{pmatrix} \\
    \tilde{\bm{\psi}}_2 &= \frac{1}{\sqrt{2}\kappa \beta}\begin{pmatrix}
        0 \\ 1
    \end{pmatrix} \ ,
\end{align}
\end{subequations}
with $\tilde{\bm{\psi}}_\alpha = \bm{0}$ for all $\alpha > 2$. Thus, only the first two modes contribute to the linear-response dissipation for the harmonic trap. From~(\ref{eq:spectralDecomposition}) the friction tensor is 
\begin{subequations}
    \begin{align}
        \tilde{\zeta} &= \beta \paren{\tau_1 \tilde{\bm{\psi}}_1\tilde{\bm{\psi}}_1^{\mathsf{T}} +  \tau_2\tilde{\bm{\psi}}_2\tilde{\bm{\psi}}_2^{\mathsf{T}}} \\
        &= \begin{pmatrix}
            \gamma & 0 \\
            0 & \frac{\gamma}{4\beta \kappa^3}
        \end{pmatrix} \ ,
    \end{align}
\end{subequations}
in agreement with \cite{sivakThermodynamicMetricsOptimal2012}. 

\subsection{Classical spin systems \label{ssec:spinSystems}}
\subsubsection{Complete control set for $N$ spins}
Control of classical spin systems in the linear-response regime has been studied fairly extensively~\cite{rotskoffOptimalControlNonequilibrium2015, rotskoffGeometricApproachOptimal2017, louwerse_connections_2022}. For a system of $N$ spins $\sigma_i$ that take values in $\curly{-1, 1}$, there are $2^N$ states $\bm{\sigma} = \paren{\sigma_0, \sigma_1, \dots, \sigma_{N-1}}$, giving rise to a $2^N$-dimensional global manifold. We begin by constructing a Riemannian full-control manifold by introducing a $(2^N -1)$-parameter complete-control set for arbitrary spin systems. Define the energy function
\begin{equation} \label{eq:clusterEnergy}
    V(\bm{\sigma}) = - \sum_{k=1}^{2^N - 1} X^k \prod_{i \in \mathcal{C}(k)} \sigma_i \ , 
\end{equation}
where $\curly{X^k}$ is the set of all $n$-spin interactions with $n = 1,\dots, N$, and $\mathcal{C}(k)$ is the set of spin indices corresponding to the coupling $X^k$. See, e.g., \eqref{eq:fourSpinEnergy} for four spins. This energy function is often called the cluster expansion~\cite{huang_finding_2016}, and it arises in the study of spin glasses and machine learning as the generalized Ising model~\cite{mezardSpinGlasses2009}, generalized lattice-gas model~\cite{decelleInferringEffectiveCouplings2024}, or generalized restricted Boltzmann machine~\cite{bulsoRestrictedBoltzmannMachines2021}. 

The probability $p(\bm{\sigma})$ of any spin configuration $\bm{\sigma}$ can be expanded in the spin moments~\cite{glauberTimeDependentStatisticsIsing1963}: 
\begin{equation} \label{eq:glauberDecomp}
    p(\bm{\sigma}) = \frac{1}{2^N} \left( 1 + \sum_j \angbr{\sigma_j}\sigma_{j} + \sum_{j\neq k}\angbr{\sigma_j\sigma_k}\sigma_{j}\sigma_{k}+\cdots \right) \ . 
\end{equation}
Following Sec.~\ref{ssec:riemFCmanifold}, we compile the forces $f_k = -\partial V/\partial X^k$ into a Jacobian matrix $\sbrac{\mathbb{J}}_{k}^{\phantom{k}\nu} = -f_k(\bm{\sigma}_{\nu})$ for some ordering of the states $\nu=0,1,\dots,2^N -1$. Then~\eqref{eq:glauberDecomp}
can be put into the form of Eq.~\eqref{eq:completenessCondition2}:
\begin{equation} \label{eq:isinginverserel}
    \bm{p} = -\frac{1}{2^N}\mathbb{J}^{\mathsf{T}}\angbr{\bm{f}} + \frac{1}{2^N}\bm{1} \ . 
\end{equation}
Thus, there is an invertible transformation between the full set of spin moments $-\angbr{\bm{f}}$ and the probability distribution $\bm{p}$. We may identify $\mathbb{J}^{\mathsf{R}} = 2^{-N}\mathbb{J}^{\mathsf{T}}$ from the above expression or derive it directly (App.~\ref{app:rightInverseIsing}). This invertibility means that the dynamics of the full set of central moments $\angbr{\delta f_k}$ is fully specified by their current values:
\begin{equation}
    \frac{\md\angbr{\delta\bm{f}(t)}}{\md t} = \mathbb{W}_{\bm{f}}\angbr{\delta\bm{f}(t)}
\end{equation}
with $\mathbb{W}_{\bm{f}} = 2^{-N}\mathbb{J}\,\mathbb{W}\,\mathbb{J}^{\mathsf{T}}$. This holds for arbitrary transition rates satisfying detailed balance.

\subsubsection{Four-spin model \label{sssec:4spin}}
We consider a four-spin system evolving under the single-spin flip transition rates
\begin{equation} \label{eq:rateLaw}
    w_{\mu\nu} = \begin{dcases} \frac{1}{1 + \exp\left\{\beta\left[V^\mu - V^{\nu}\right]\right\}} \ , & \text{$\bm{\sigma}_\mu,\bm{\sigma}_\nu$ differ in one spin} \\
    0 \ , & \text{otherwise} 
    \end{dcases}
\end{equation}
for $\mu \neq \nu$. These are well-studied dynamics for the Ising system, sometimes called heat-bath dynamics or Glauber dynamics~\cite{marizMixingHeatbathGlauber1989, aurellCavityMasterEquation2017, marizGeneralizedSinglespinflipDynamics1994, louwerse_connections_2022}. 

\begin{figure}
    \centering
    \includegraphics[width=246 pt]{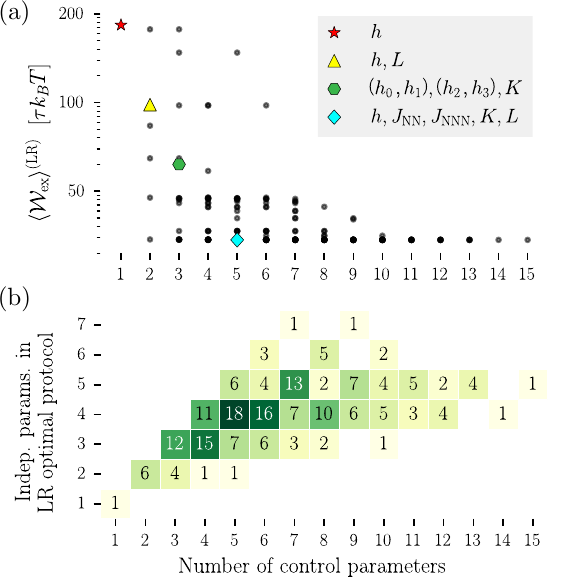}
    \caption{(a) LR excess work for LR optimal protocols for the four-spin system computed for 200 different control sets with between $1$ and $15$ control parameters. The protocols featured in Fig.~\ref{fig:threeProtocols} and Fig.~\ref{fig:eigPlot} are labeled. (b) Two-dimensional histogram showing the number of independently varying parameters against the number of available control parameters (e.g., 18 of the calculated 5-parameter optimal protocols had only 4 independently varying parameters).}
    \label{fig:isingall}
\end{figure}

For a system of four spins, the generalized Ising model~(\ref{eq:clusterEnergy}) has 15 parameters: 4 fields $h_i$, 6 two-spin couplings $J_{ij}$, 4 three-spin interactions $K_{ijk}$, and a single four-spin interaction $L$ (Fig.~\ref{fig:fourspinParams}, App.~\ref{app:cSets}): 
\begin{align}\label{eq:fourSpinEnergy}
    V(\bm \sigma) = - \sum_i &h_i\sigma_i -\sum_{i < j}J_{ij}\sigma_i\sigma_j \nonumber  \\
    &-\sum_{i < j <k}K_{ijk}\sigma_i\sigma_j\sigma_k - L \prod_i\sigma_i
\end{align}
From these 15 parameters, the imposition of constraints on their time evolution may be used to build partial control sets. \hyperref[fig:modelSystems]{Figure 4(c)} shows one such partial control set, with a bulk field $h =h_0=h_1=h_2=h_3$, two-spin coupling blocks $J_{\text{NN}}=J_{01}=J_{02}=J_{13}=J_{23}$ (for nearest neighbors) and $J_{\text{NNN}}=J_{03}=J_{12}$ (for next-nearest neighbors), bulk three-spin coupling $K=K_{012}=K_{013}=K_{023}=K_{123}$, and four-spin coupling $L$. 

We set our boundary conditions in terms of these 5 bulk couplings to $h(0)=-1, J_{\text{NN}}(0)=1$ and $h(\tau) = +1, J_{\text{NN}}(\tau)=1$, with all other couplings set to zero at $t =0$ and $t=\tau$. Though \eqref{eq:fourSpinEnergy} provides no notion of ``geometry'' for the spin system, under these boundary conditions it is constructive to imagine a ferromagnetic Ising model on a $2 \times 2$ grid as in \hyperref[fig:modelSystems]{Fig.~4(c)}.    

\begin{figure*}
    \centering
    \includegraphics[width=7.06 in]{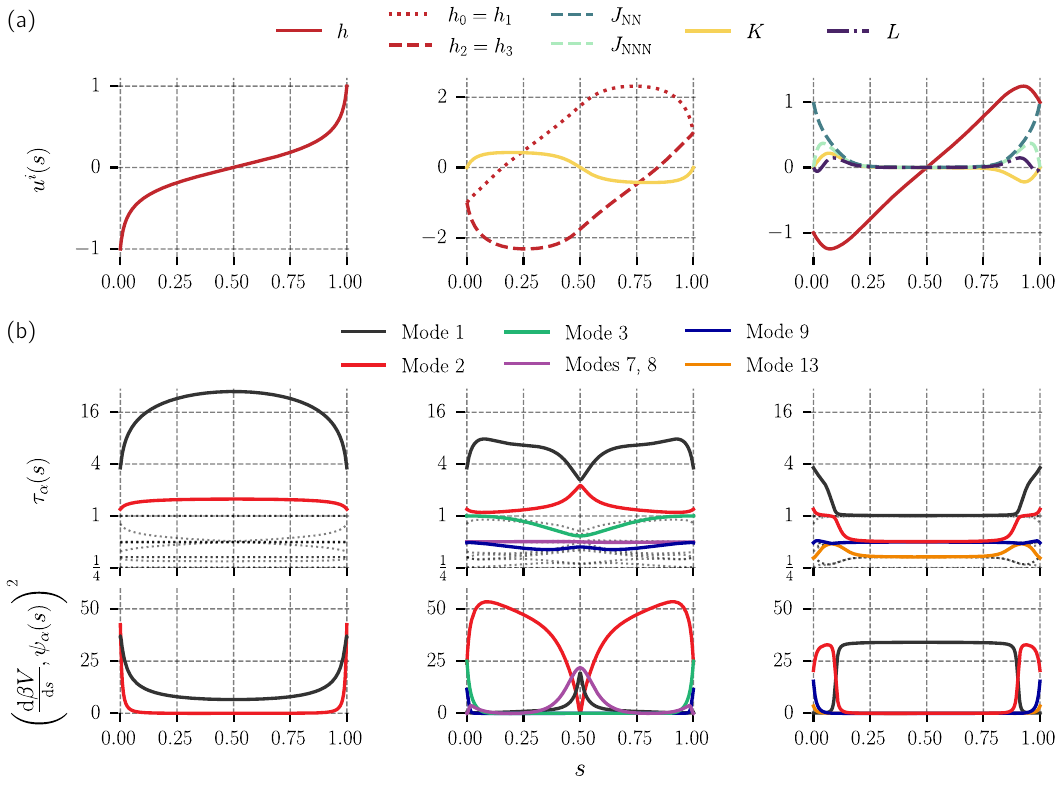}
    \caption{
    (a) LR optimal control protocols for a classical four-spin system in scaled time $s = t/\tau$, for three different control-parameter sets. (b) Spectral contributions to the LR excess power~(\ref{eq:spectralDecomposition}). The dissipation along mode $\alpha$ is the product of the relaxation time $\tau_{\alpha}(s)$ and the squared inner product $\paren{\md \beta V /\md s, \psi_{\alpha}}^2$ of the control-parameter velocity with the instantaneous mode. Modes are ordered such that $\tau_1(0) \geq \tau_2(0) \geq \cdots$. In the middle panel, the squared inner products of the degenerate modes $7$ and $8$ with the velocity are summed since their respective contributions to the dissipation cannot meaningfully be distinguished. Dotted black lines: modes that are everywhere orthogonal (or nearly so) to the velocity $\md \beta V/\md s$ and which thus do not meaningfully contribute to dissipation. 
    }
    \label{fig:threeProtocols}
\end{figure*}

We computed LR minimum-work protocols for 200 different control sets consistent with these endpoints. The number of control parameters per set varied from $1$ for the minimal set (with only a global field $h$) to $15$ for the maximal set (with independent control over all parameters in the cluster expansion). Appendix~\ref{app:cSets} details the construction of control sets and computation of minimum-work protocols. 

The results are summarized in Fig.~\ref{fig:isingall}. We observe that very few protocols need to use all available degrees of freedom to minimize the linear-response excess work. For instance, the optimized 15-parameter full-control protocol is (to numerical precision) identical to the 5-parameter protocol [\hyperref[fig:isingall]{Fig.~5(a)} blue diamond, and Fig.~\ref{fig:threeProtocols} right]. Evidently (at least within the linear-response regime) it is not possible to reduce dissipation by assuming a finer level of control over the spin system. In all, only 42 distinct protocols were identified among the 200 computed. Figure~\ref{fig:isingall}b shows the number of independently varying parameters compared to the number of available parameters; at most 7 parameters were utilized. 

Figure~\ref{fig:threeProtocols} shows LR-optimized protocols for 3 different sets of control parameters: the one-parameter set $(h, \neg J, \neg K, \neg L )$, the three-parameter set $(h_{=}, \neg J, K, \neg L)$ and the five-parameter set $(h, (J_{\text{NN}}, J_{\text{NNN}}), K, L)$. Here the four elements of the sets refer to constraints on control of the $1$-, $2$-, $3$-, and $4$-spin parameters, respectively (see Table~\ref{tab:csets} of App.~\ref{app:cSets}). For instance, $h_{=}$ represents two wide magnetic fields, one interacting with the upper row of spins ($\sigma_0,\sigma_1$) and the other with the bottom row ($\sigma_2,\sigma_3$), realized by constraints $h_0 = h_1$ and $h_2 = h_3$. The symbol $\neg$ indicates no control of the given type of coupling, so $\neg J$ indicates that $J_{\text{NN}} = 1$ and $J_{\text{NNN}}=0$ are fixed.

The one-parameter protocol (Fig.~\ref{fig:threeProtocols}, left) controls only a global field $h$. This is the minimal control set for the specified protocol endpoints. Since the region of control-parameter space traversed by $h$ is fixed, the only optimizable property is the velocity of $h(t)$, which for the optimal protocol slows down where relaxation times are relatively large. The dominant contribution by far to the dissipation comes from the first mode.

The other two protocols [\hyperref[fig:threeProtocols]{Fig.~6(a)}, middle and right] are geodesics on three- and five-dimensional thermodynamic manifolds. Both protocols change the magnetic fields nonmonotonically, a feature previously observed in Ref.~\cite{louwerseMultidimensionalMinimumworkControl2022}. How this (perhaps surprising) behavior reduces the total excess work of the protocols is revealed by examining the separate spectral contributions to the LR excess power~\eqref{eq:spectralPower}.  

\begin{figure}
    \centering
    \includegraphics[width=246 pt]{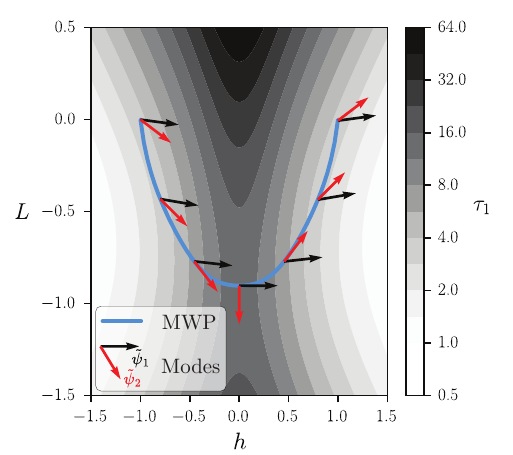}
    \caption{Projections $\tilde{\bm{\psi}}_1$ and $\tilde{\bm{\psi}}_2$ of the respective first and second modes (black and red arrows) onto the thermodynamic manifold for the $(h,L)$ control set plotted along the linear-response MWP (blue curve). For visual clarity, the vectors are plotted with unit length, although their magnitudes vary across the space. Heatmap: relaxation time $\tau_1$ of the first mode $\bm\psi_1$.}
    \label{fig:eigPlot}
\end{figure}

The first mode $\bm{\psi}_1(s)$ has a fairly persistent identity in the explored regions of parameter space: roughly, it characterizes relaxation between the two ferromagnetic states (see also Fig.~\ref{fig:eigPlot}). Given that the central task is to flip the spins from spin-down to spin-up on average, it is not surprising that the first mode was the dominant contribution to the LR excess work for almost all of the 200 parameter sets examined. 

Projecting the state energies onto the relaxation modes (Fig.~\ref{fig:threeProtocols}, bottom row) reveals that driving is nearly orthogonal to the first mode along the nonmonotonic portions of the protocols, and thus concentrates nearly all dissipation among modes other than the first during these time windows. For the 5-parameter protocol, this feature significantly reduces the relaxation time $\tau_1$. This uncovers the control strategy: by driving orthogonally to $\bm{\psi}_1$, we are able to efficiently drive the system to a region of control-parameter space in which $\tau_1$ is relatively small. 

For the 3-parameter protocol, not only is $\tau_1$ \textit{not} significantly reduced, it is roughly doubled near the ends of the protocol. In breaking the symmetry of the four spins, the driving is dominantly directed along the relatively fast-relaxing second mode $\bm{\psi}_2$, which can be interpreted as relaxation between the ferromagnetic states and the state $\bm{\sigma} = (+1,+1,-1,-1)$. We also observe that (somewhat counterintuitively) most of the $\sim$15\% of dissipation attributable to the first mode occurs near $s = 1/2$ (when the spectral gap is relatively small), while most of the $\sim$80\% of dissipation attributable to the second mode occurs when the spectral gap is largest. While $\tau_1$ is often regarded as \textit{the} characteristic dynamical times for a system with a large spectral gap~\cite{blaber_efficient_2022, roux_transition_2022}, our results show that in driven systems faster modes can become salient, and may even dominate the dissipation -- especially for efficient protocols which may drive orthogonal to slow modes to save energy. 

Finally, Fig.~\ref{fig:eigPlot} shows the linear-response minimum-work protocol for the two-parameter control set $(h,\neg J, \neg K, L)$. The control strategy is similar to the 5-parameter MWP: in the relevant region of the $(h,L)$-manifold, the projection $\tilde{\bm{\psi}}_1$ of the first mode (black arrows) is nearly parallel to the $h$-direction. So, given that one \textit{must} drive along the first mode in order to flip the spins, reduced dissipation is achieved by driving the system nearly perpendicularly to this mode to reach a region of control-parameter space where $\tau_1$ is smaller.

\section{Summary and Outlook}
We have presented a foundation for the friction-tensor formalism for conservative control of stochastic systems. Thermodynamic manifolds $\tilde{\mathcal{M}}$ for partial control sets $\bm{u}$ are derived as submanifolds of a global thermodynamic manifold $\mathcal{M}$ whose metric structure is provided by the full-control friction tensor. We derived two decompositions of the full-control friction tensor which facilitate both interpretation and calculation of arbitrary friction tensors and minimum-work protocols. 

In particular, we show that the linear-response contribution to the excess power can be decomposed into a sum of terms corresponding to the system's relaxation modes~(\ref{eq:spectralPower}). The magnitude of the $\alpha$th term depends on the degree to which the protocol drives the system along that mode, scaled by its relaxation time $\tau_\alpha$. This decomposition suggests a design principle for protocols that reduce excess dissipation for finite-time processes: one should drive a system such that $\dd{V}/\dd{t}$ is maximally orthogonal to modes with long relaxation times. When this is possible, the common identification of $\tau_1$ as the system's characteristic dynamical timescale may be unsuitable even in the presence of a large spectral gap.

For low-dimensional discrete systems with small state spaces or continuous systems with known spectra, these decompositions simplify the analytic calculation of the friction tensor. For larger systems, the operator decomposition (\ref{eq:dynamicalDecompositionDiscrete}) enables numerical calculation of the friction tensor without recourse to simulation of the stochastic dynamics. Indeed, for the 4-spin system with $|\Omega| = 16$, we were able to rapidly compute 200 different MWPs---many on high-dimensional thermodynamic manifolds, with up to 15 control parameters. This opens up the possibility of systematically searching control spaces for new insights and principles.

For very large discrete systems or continuous systems discretized on a fine grid, the matrix inversion in~(\ref{eq:dynamicalDecompositionDiscrete}) will be prohibitively expensive. Truncation of the spectral decomposition may enable controlled-accuracy approximation of the friction tensor, particularly in the case of a large spectral gap (i.e., $\tau_1 \geq \cdots \geq \tau_q \gg \tau_{q+1} \geq \cdots$ for some $q$). However, given that even long relaxation timescales may be irrelevant when driving is orthogonal to the corresponding mode, the accuracy of this approximation fundamentally depends on the nature of driving.

A natural extension concerns higher-order corrections to the excess work. From the relation~\eqref{eq:LRrelation} and the operator decomposition~\eqref{eq:dynamicalDecomposition}, one may iteratively construct an asymptotic series of the excess work in powers of $\tau^{-1}$, as shown in~\cite{mandalAnalysisSlowTransitions2016} for discrete systems. Future research on this subject could address the convergence properties of the formal series and optimal truncation.

\begin{acknowledgments}
We thank Adrianne Zhong (University of California Berkeley, Department of Physics) and Antonio Patr\'on Castro and W.\ Callum Wareham (Simon Fraser University, Department of Physics) for insightful conversations and feedback on the manuscript. This work was supported by a Natural Sciences and Engineering Research Council of Canada (NSERC) Undergraduate Student Research Award (J.R.S.), NSERC CGS Master's and Doctoral scholarships (J.R.S.), an NSERC Discovery Grant RGPIN-2020-04950 (D.A.S.), and a Tier-II Canada Research Chair CRC-2020-00098 (D.A.S.), and was enabled in part by support provided by the Digital Research Alliance of Canada (alliancecan.ca). 
\end{acknowledgments}

\section*{Data availability}
The data that support the findings of this article are openly available \cite{sawchuk_2025_15678284}. 

\appendix 

\section{Completeness condition \label{app:fcc}}
A set of control parameters $\bm{u}$ is \textit{complete} when it can manipulate the energy landscape $\bm{V}(\bm{u})$ up to a global additive constant. For $|\Omega| = n$, this requires $n-1$ parameters. We show that when the hypersurface $\bm{V}(\bm{u})$ is perpendicular to the null direction $\bm{1}$, the completeness condition Eq.~\eqref{eq:completenessFormal} is equivalent to the statement that $\angbr{\bm{f}}_{\bm{p}(t)} = - \mathbb{J}\bm{p}(t)$ has inverse
\begin{equation} \label{eq:app1_2}
    \bm{p}(t) = - \mathbb{J}^{\mathsf{R}}\angbr{\bm{f}(t)} + \frac{1}{n}\bm{1} \ 
\end{equation}
for Jacobian $\mathbb{J}_i^{\phantom{i}\nu} = \partial V^\nu/\partial u^i$ and right inverse $\mathbb{J}^{\mathsf{R}} \equiv \mathbb{J}^{\mathsf{T}}(\mathbb{J}\mathbb{J}^{\mathsf{T}})^{-1}$.

First we show that \eqref{eq:completenessFormal}$\implies$\eqref{eq:app1_2}. The condition~\eqref{eq:completenessFormal} implies that $\mathbb{J}$ has full row rank, so the right inverse of the Jacobian for a complete control set exists, and satisfies 
\begin{subequations} 
\begin{align}
    \mathbb{J}\mathbb{J}^{\mathsf{R}} &= I_{n-1} \label{eq:rightOnRight} \\
    \mathbb{J}^{\mathsf{R}} \mathbb{J} &= I_n - \frac{1}{n}\mathbbm{1}_n \ , \label{eq:rightOnLeft}
\end{align}
\end{subequations}
where $I_n$ is the $n\times n$ identity matrix and $\mathbbm{1}_n$ is the $n\times n$ matrix of ones. Equation~\eqref{eq:rightOnLeft} follows from the facts that $\mathbb{J}^{\mathsf{R}} \mathbb{J}$ is the transpose of the orthogonal projector onto the span of $\mathbb{J}$~\cite{yanaiProjectionMatricesGeneralized2011} and the null-space $\mathcal{N}(\mathbb{J})$ is spanned by the vector $\bm{1}$. 

Operating on Eq.~\eqref{eq:avgFJac} with $\mathbb{J}^{\mathsf{R}}$ and applying the identity \eqref{eq:rightOnLeft} yields
\begin{equation} \label{eq:rightRel}
    \mathbb{J}^{\mathsf{R}}\angbr{\bm{f}(t)} = -\bm{p}(t) + \frac{1}{n}\bm{1} \ ,
\end{equation}
so Eq.~\eqref{eq:app1_2} holds for any complete control set $\bm{u}$. 

To prove the converse (i.e., \eqref{eq:app1_2}$\implies$\eqref{eq:completenessFormal}), the existence of a right inverse $\mathbb{J}^{\mathsf{R}}$ immediately implies that $\mathcal{N}(\mathbb{J})$ is one-dimensional. Left-multiplying~\eqref{eq:rightRel} by $\mathbb{J}$ and applying \eqref{eq:rightOnRight} gives
\begin{subequations}
    \begin{align}
         \angbr{\bm{f}(t)}&= -\mathbb{J}\bm{p}(t) + \frac{1}{n}\mathbb{J}\bm{1} \\
         &= \angbr{\bm{f}(t)} + \frac{1}{n}\mathbb{J}\bm{1} \ ,
    \end{align}
\end{subequations}
so $\mathcal{N}(\mathbb{J}) = \text{span}\curly{\bm{1}}$.

\section{Equilibrium covariance matrix of the indicator functions \label{app:eqcov}}
We derive here a simple expression for the equal-time equilibrium covariance matrix $\Sigma_{\omega}^0$ of the forces $\omega(x,t)$ conjugate to the potential energy $V(x)$:
\begin{subequations}
    \begin{align}
        \Sigma^0_{\omega}(x,y) &\equiv \angbr{\delta \omega(x)\delta\omega(y)} \\
        &= \angbr{\omega(x)\omega(y)} - \pi(x)\pi(y) \\
        &= \int_{\Omega}\md\mu(z)\,\pi(z)\delta(x,z)\delta(y,z) - \pi(x)\pi(y) \\
        &= \pi(x)\delta(x,y) - \pi(x)\pi(y) \ .
    \end{align}
\end{subequations}
For discrete-state systems, it is useful to write this in matrix form:
\begin{align}\label{eq:covd}
    \phantom{\angbr{\delta\omega \, \delta}}\Sigma^0_{\omega} &= \text{diag}\curly{\bm{\pi}} - \bm{\pi}\bm{\pi}^{\mathsf{T}} \ .
\end{align}

\section{Riemannian full-control friction tensor\label{app:fullControlFriction}}
We show that the friction tensor for a complete control set is given by Eq.~(\ref{eq:completeControlFriction}). First, note that from the definition (\ref{eq:generatorF}) of the generator for the dynamics of $\angbr{\bm{f}(t)}$ and the properties (\ref{eq:rightOnRight}) and (\ref{eq:rightOnLeft}) for the right inverse $\mathbb{J}^R$, the inverse $\mathbb{W}_{\bm{f}}^{-1}$ exists and equals
\begin{equation}
    \mathbb{W}_{\bm{f}}^{-1} = \mathbb{J}\mathbb{W}^{\mathcal{D}}\mathbb{J}^{\mathsf{R}} \ .
\end{equation}
This is straightforwardly verified using the definitions and properties of $\mathbb{W}_{\bm{f}}, \mathbb{W}^{\mathcal{D}}$, and $\mathbb{J}^{\mathsf{R}}$. Next, 
\begin{subequations}
    \begin{align}
        \sbrac{\mathbb{J}\Sigma^0_{\omega}\mathbb{J}^{\mathsf{T}}}_{ij} &= \sum_{\mu,\nu}f_{i}(x_{\mu})f_j(x_{\nu})\paren{\pi_{\mu}\delta_{\mu\nu} - \pi_{\mu}\pi_{\nu}} \\
        &= \angbr{f_if_j}_{\pi} - \angbr{f_i}_{\pi}\angbr{f_j}_{\pi} \\
        &= \sbrac{\Sigma^0_{\bm{f}}}_{ij} \ ,
    \end{align}
\end{subequations}
where in the first line we used Eq.~(\ref{eq:covd}). Then 
\begin{subequations}
    \begin{align}
        -\beta{\mathbb{W}}_{\bm{f}}^{-1}\Sigma^0_{\bm{f}} &= -\beta\mathbb{J}\mathbb{W}^{\mathcal{D}}\Sigma^0_{\omega}\mathbb{J}^{\mathsf{T}} \\
        &= \mathbb{J}\zeta\mathbb{J}^{\mathsf{T}} \\ 
        &= \tilde{\zeta} \ .
    \end{align}
\end{subequations}

\section{Spectral decomposition of the Drazin inverse \label{app:specDecDraz}}
Let $\mathcal{G}$ be the generator of a discrete- or continuous-state Markov process with spectral decomposition 
\begin{equation}
    \mathcal{G}(x,y) = \sum_{\alpha \geq 1} \lambda_{\alpha}\psi_{\alpha}(x)\phi_{\alpha}(y) \ .
\end{equation}
We will show that the integral kernel for the Drazin inverse of $\mathcal{G}$ is
\begin{equation} \label{eq:generalDrazSpec}
    \mathcal{G}^{\mathcal{D}}(x,y) = -\sum_{\alpha \geq 1}\frac{1}{\lambda_{\alpha}}\psi_{\alpha}(x)\phi_{\alpha}(y) \ .
\end{equation}
First, the composition of $\mathcal{G}$ with $\mathcal{G}^{\mathcal{D}}$ is
\begin{subequations}
    \begin{align}
        (&\mathcal{G}\circ \mathcal{G}^{\mathcal{D}})(x,y) \nonumber \\
        &= \int_{\Omega}\md \mu(z) \, \mathcal{G}(x,z) \, \mathcal{G}^{\mathcal{D}}(z,y) \\
        &= \sum_{\alpha,\beta \geq 1}\frac{\lambda_{\alpha}}{\lambda_{\beta}}\psi_\alpha(x)\phi_\beta(y)\int_{\Omega}\md\mu(z) \, \phi_\alpha(z)\psi_\beta(z) \\
        &= \sum_{\alpha \geq 1} \psi_{\alpha}(x)\phi_{\alpha}(y) 
    \end{align}
\end{subequations}
where the last line follows from the biorthogonality of the eigenfunctions. The composition $\mathcal{G}^{\mathcal{D}}\circ\mathcal{G}$ yields the same result, so the commutativity property (\ref{eq:drazinprop3}) is satisfied.

Since the full set of eigenfunctions is also complete, i.e.,
\begin{equation}
    \sum_{\alpha \geq 0}\psi_\alpha(x)\phi_{\alpha}(y) = \delta(x,y) \ ,
\end{equation}
this further simplifies to 
\begin{equation} \label{eq:appDrazProjector}
    (\mathcal{G}\circ \, \mathcal{G}^{\mathcal{D}})(x,y) = \delta(x,y) - \psi_0(x) \ .
\end{equation}
This is equivalent to Eq.~(\ref{eq:drazinProjection}). That is, $\mathcal{G}\circ\mathcal{G}^{\mathcal{D}}$ projects vectors in the Hilbert space onto the complement of $\mathcal{N}(\mathcal{G})$.

To see that properties (\ref{eq:drazinprops}a,b) hold with index $k = 1$, note that $1 \mathcal{G}^{\mathcal{D}} = \mathcal{G}^{\mathcal{D}}\psi_0 = 0$ by construction. Then 
\begin{subequations}
    \begin{align}
        (\mathcal{G}^{\mathcal{D}}\circ\mathcal{G}&\circ\mathcal{G}^{\mathcal{D}})(x,y) \nonumber \\
        &= \int \md\mu(z)\, [\delta(x,z)-\psi_0(x)] \, \mathcal{G}^{\mathcal{D}}(z,y) \\
        &= \mathcal{G}^{\mathcal{D}}(x,y) \ ,
    \end{align}
\end{subequations}
and similarly,
\begin{subequations}
    \begin{align}
        (\mathcal{G}\circ\mathcal{G}^{\mathcal{D}}&\circ\mathcal{G})(x,y) \nonumber \\ 
        &= \int \md\mu(z) \, \mathcal{G}(x,z) \, [\delta(z,y)-\psi_0(z)] \\
        &= \mathcal{G}(x,y) \ .
    \end{align}
\end{subequations}
Since the Drazin inverse is the unique operator satisfying properties~\hyperref[eq:drazinprops]{(36a)--(36c)}, $\mathcal{G}^{\mathcal{D}}$ is the Drazin inverse of $\mathcal{G}$.

\section{Proof of Eq.~\eqref{eq:fricDrazRelation} \label{app:fricDrazProof}}
To show that \eqref{eq:fricDrazRelation} holds, we first note that the temporal derivative of the instantaneous equilibrium distribution $\pi(x,t) = Z^{-1}\exp\curly{-\beta V(x,t)}$ can be expressed in terms of $\md V/\md t$ by 
\begin{equation}
\frac{\md\pi(x,t)}{\md t} 
= \int_{\Omega}\md\mu(y) \, \frac{\md V(y,t)}{\md t}\frac{\delta \pi(x,t)}{\delta V(y,t)} \ ,
\end{equation}
where $\delta \pi(x,t)/\delta V(y,t)$ is the functional derivative of $\pi$ with respect to $V$. From the definition of the Boltzmann distribution,
\begin{subequations}
    \begin{align}
        \frac{\delta \e{-\beta V(x,t)}}{\delta V(y,t)} &= -\beta \, \delta(x,y) \, \e{-\beta V(x,t)} \label{eq:appDa} \\ 
        \frac{\delta Z[V(t)]}{\delta V(y,t)} 
        &= -\beta \int_{\Omega}\md\mu(x) \, \delta(x-y) \, \e{-\beta V(x,t)} \\
        &= -\beta \e{-\beta V(y,t)} \ . \label{eq:appDb}
    \end{align} 
\end{subequations}
Then by the chain and product rules, 
\begin{subequations}
    \begin{align}
        \frac{\delta \pi(x,t)}{\delta V(y,t)} &= - \frac{1}{Z^2}\frac{\delta Z [V(t)]}{\delta V(y,t)} \, \e{-\beta V(x,t)} +\frac{1}{Z}\frac{\delta \e{-\beta V(x,t)}}{\delta V(y,t)}  \\
        &= \beta\frac{\e{-\beta V(y,t)}}{Z}\frac{\e{-\beta V(x,t)}}{Z} - \beta\delta(x,y)\frac{\e{-\beta V(x,t)}}{Z} \\
        &= \beta \, \pi(x,t)\sbrac{\pi(y,t) - \delta(x,y)}
    \end{align}
\end{subequations}
The temporal derivative of the equilibrium distribution is therefore
\begin{equation} \label{eq:refDm}
    \frac{\md\pi(x,t)}{\md t} = \beta  \,\pi(x,t)
    \left[
    \angbr{\frac{\md V}{\md t}}_{\pi(t)} -\frac{\md V(x,t)}{\md t} 
    \right]  
    \ .
\end{equation}
Since $\pi(t)$ is in the null-space of $\mathcal{G}^{\mathcal{D}}_t$, operating on both sides of \eqref{eq:refDm} 
with $\mathcal{G}^{\mathcal{D}}_t$ yields
\begin{subequations}
    \begin{align}
        \bigg(\mathcal{G}^{\mathcal{D}}_t&\frac{\md\pi}{\md t}\bigg)(x) \nonumber \\ 
        &= \int_{\Omega}\md \mu(y) \, \mathcal{G}^{\mathcal{D}}_t(x,y)\frac{\md \pi(y)}{\md t} \\
        &= \sum_{\alpha \geq 1}\tau_{\alpha}\psi_{\alpha}(x)\int_{\Omega}\md\mu(y) \, \phi_{\alpha}(y)\pi(y)\frac{\md V(y)}{\md t} \label{eq:spectralstep} \\
        &= \sum_{\alpha \geq 1}\tau_{\alpha}\psi_{\alpha}(x)\int_{\Omega}\md\mu(y) \, \psi_{\alpha}(y)\frac{\md V(y)}{\md t} \\
        &= \paren{\zeta \frac{\md V}{\md t}}(x) \ ,
    \end{align}
\end{subequations}
where in \eqref{eq:spectralstep} we applied the spectral decomposition~\eqref{eq:generalDrazSpec} of the Drazin inverse of the generator.

\section{Right inverse of the Ising full-control Jacobian \label{app:rightInverseIsing}}
We show explicitly that the Jacobian matrix $\mathbb{J}$ for the the full set of control parameters from the cluster expansion~(\ref{eq:clusterEnergy}) has right inverse $\mathbb{J}^{\mathsf{R}} =2^{-N}\mathbb{J}^{\mathsf{T}}$, as inferred from \eqref{eq:isinginverserel}. The components of $\mathbb{J}$ are 
\begin{equation}
    \sbrac{\mathbb{J}}_{k}^{\phantom{i}\nu} = -\prod_{k \, \in \, \mathcal{C}(k)}\sigma_j(x_{\nu}) \ ,
\end{equation}
where $\mathcal{C}(k)$ is the set of spin indices associated with interaction $X^k$, and $\sigma_j(x_{\alpha})$ is the value of spin $k$ in state $x_\alpha$. Then
\begin{subequations}
    \begin{align}
        \sbrac{\mathbb{J}\mathbb{J}^{\mathsf{T}}}_{k\ell}
        &= \sum_{\nu}\prod_{i  \in  \mathcal{C}(k)}\prod_{j \in  \mathcal{C}(\ell)}\sigma_i(x_{\nu}) \, \sigma_j(x_{\nu}) \ .
    \end{align}
\end{subequations}

For $k=\ell$, the spin sets are identical so the summand evaluates to one and thus $\sbrac{\mathbb{J}\mathbb{J}^{\mathsf{T}}}_{kk} = 2^N$. 

For $k\neq \ell$, we need only consider spin indices in \textit{either} $\mathcal{C}(k)$ \textit{or} $\mathcal{C}(\ell)$, since any $i \in \mathcal{C}(k)\cap\mathcal{C}(\ell)$ contributes a factor of one to the product. Denote this set by $\mathcal{C}(k)\oplus\mathcal{C}(\ell)$. For each state $x_\alpha$, there is a state $x_{\overline{\alpha}}$ obtained by inverting all spins. If $\mathcal{C}(k)\oplus\mathcal{C}(\ell)$ has an \textit{odd} number of spins, then the $\alpha$th and $\overline{\alpha}$th terms in the sum cancel, so these entries will vanish. If $\mathcal{C}(k)\oplus\mathcal{C}(\ell)$ has an \textit{even} number of spins, a similar cancellation occurs by inverting only the first spin in each (ordered) product. Therefore, $\sbrac{\mathbb{J}\mathbb{J}^{\mathsf{T}}}_{k\ell} = 2^N \delta_{k\ell}$, so 
\begin{equation}
    \mathbb{J}^{\mathsf{R}} 
    = \frac{1}{2^N}\mathbb{J}^{\mathsf{T}} \ .
\end{equation}

\section{Control sets and numerical methods for the four-spin system \label{app:cSets}}
\begin{figure}[H]
    \centering
    \includegraphics[width=246 pt]{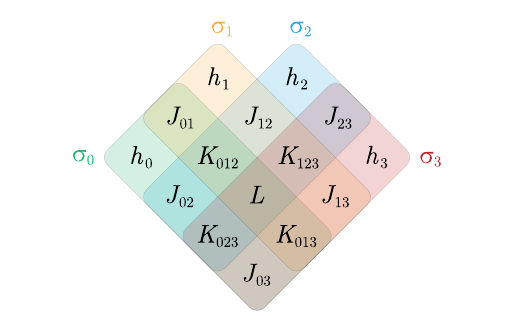}
    \caption{Fourth-order Venn diagram showing the 15 control parameters in a complete control set for the four-spin model. Each spin is associated with a set of 8 different parameters, and the 4-spin interaction $L$ is at the intersection of all four sets.}   
    \label{fig:fourspinParams}
\end{figure}

We computed minimum-work protocols for a collection of 200 control-parameter sets with between $1$ and $15$ parameters for the four-spin system. The endpoints for all protocols are $h(0)=-1$ and $h(\tau)=+1$ for global field $h \ (h_0 = h_1 = h_2 = h_3)$, $J_{\text{NN}}(0)=J_{\text{NN}}(\tau)=1$ for ``nearest-neighbor'' coupling $J_{\text{NN}} \ (J_{01}=J_{02}=J_{13}=J_{23})$, and all other interactions set to zero. Any control set capable of driving the system between these endpoints must have (at minimum) control over all of the fields $h_i$. 

The parameter sets studied are the elements of the product set
\begin{align} \label{eq:controlSets}
    &\sbrac{h, h_{=}, h_{\times}, h_i} \nonumber \\
    & \ \ \ \ \ \times \sbrac{\neg J, J_{\text{NN}}, J_{\text{NNN}}, (J_{\text{NN}}, J_{\text{NNN}}), J_{ij}} \nonumber  \\
    & \ \ \ \ \ \times \sbrac{\neg K, K, K_{=}, K_{\times}, K_{ijk}} \nonumber \\
    & \ \ \ \ \ \times \sbrac{\neg L, L} \ .
\end{align}
The elements in the sets above are control subsets defined by constraints on the $1$-, $2$-, $3$-, and $4$-spin interactions. For example, referring to Table \ref{tab:csets}, the subset $h_{=}$ contains two parameters, $J_{\text{NNN}}$ and $L$ each contain one parameter, and $\neg K$ is an empty set, so the control-parameter set $(h_{=},J_{\text{NNN}},\neg K,L$) contains the four parameters
\begin{equation}
    \begin{matrix}
        u^0 &= &h_0= h_1 \\ u^1 &= &h_2 = h_3 \\ u^2 &= &J_{03} = J_{12} \\ u^3 &= &L \ .
    \end{matrix}
\end{equation}

\begin{table}[H]
    \centering
    \begingroup
    \renewcommand{\arraystretch}{1.2}
    \begin{tblr}{colspec = {Q[c,m] | Q[c,m]},
             row{odd[0]} = {bg=gray!15}
             }
        \SetRow{}
         $h$ & $(h_0= h_1= h_2= h_3)$  \\ 
         $h_{=}$    &  $(h_0= h_1),(h_2=h_3)$ \\
         $h_{\times}$ & $(h_0= h_3), (h_1=h_2)$ \\
         $h_i$ & All fields independently controlled \\
         $\neg J$ & $J_{\text{NN}}$=1, $J_{\text{NNN}}$=0 (uncontrolled)   \\
         $J_{\text{NN}}$ & $(J_{01}= J_{02}= J_{13}= J_{23})$ \\
         $J_{\text{NNN}}$ & $(J_{03}= J_{12})$ \\
         $J_{ij}$ & All two-spin couplings independently controlled \\
         $\neg K$ & $K = 0$ (uncontrolled)  \\
         $K$ & $(K_{012}= K_{013}= K_{023}= K_{123})$ \\
         $K_{=}$ & $(K_{012}= K_{013}), (K_{023}= K_{123})$ \\
         $K_{\times}$ & $(K_{012}= K_{023}), (K_{013}= K_{123})$ \\
         $K_{ijk}$ & All three-spin interactions independently controlled \\
         $\neg L$ & $L = 0$ (uncontrolled) \\
         $L$ & Four-spin interaction controlled 
    \end{tblr}
    \endgroup
    \caption{Defining constraints of the control subsets comprising the 200 parameter sets in~\eqref{eq:controlSets}.}
    \label{tab:csets}
\end{table}

To calculate the MWPs, we used a relaxation method~\cite{rotskoffGeometricApproachOptimal2017, louwerseMultidimensionalMinimumworkControl2022}. An auxiliary variable $r$ is introduced, defining a system of partial differential equations
\begin{subequations} 
    \begin{empheq}[left=\empheqlbrace]{align}
        & \frac{\partial^2 u^i}{\partial t^2} + \sum_{j,k}\Gamma^i_{\phantom{i}jk}(\bm{u})\frac{\partial u^j}{\partial t}\frac{\partial u^k}{\partial t} = -\frac{\partial u^i}{\partial r} \label{eq:PDE} \\
        & u^i(0,r)= u^i_0, \ u^i(\tau, r) = u^i_{\tau} \ \ \ \forall \, r \label{eq:endpoints}\\
        & u^i(t,0) = g^i(t) \label{eq:initialization}
    \end{empheq}
\end{subequations} ~\\
for protocol endpoints $u^i_0$ and $u^i_{\tau}$ and initial curve $g^i(t)$. If this system is sufficiently well-behaved, the curve $u^i(t,r)$ approaches a solution of the geodesic equation~(\ref{eq:geodesicequation}) as $r \to \infty$ for any initialization $g^i(t)$. When $r$ and $t$ are discretized, $\Delta r$ plays the role of a learning rate, scaling the magnitude of the updates $u^i(t,r) \to u^i(t, r+\Delta r)$. We took $\Delta r \sim \Delta t^2$ to ensure stability. Rapid convergence was achieved by initializing with a large timestep $\Delta t = 0.02$ and progressively refining the grid to a maximum resolution $\Delta t = 0.00125$. 

We obtained an analytic (if inelegant) expression for the derivatives of the friction tensor using the operator decomposition~(\ref{eq:dynamicalDecompositionDiscrete}) with the rate law $w_{\mu\nu} = \pi_\mu / (\pi_\mu + \pi_\nu)$ for Boltzmann distribution $\bm \pi$ and the form~(\ref{eq:drazininverse}) of the Drazin inverse. The Christoffel symbols~(\ref{eq:christoffeltwo}) were then rapidly computed on-the-fly from local information at each point on the discretized curve. 

The spectra exhibit eigenvalue-crossings, so ordinary rank-ordering (i.e., $\lambda_0(t) \geq \lambda_1(t)\geq \cdots$ for all $t$) results in non-smooth evolution $\lambda_\alpha(t)$ of the eigenvalues and discontinuities in the evolution $\bm{\psi}_{\alpha}(t)$ of eigenvectors.  In order to preserve continuity of physically meaningful eigenvectors, we computed the rank-ordered spectrum at $t = 0$ and used Rayleigh-quotient iteration~\cite{trefethenNumLinAlg} over a fine interpolation of the protocols ($\Delta t = 8.33\times10^{-5}$, or 12,000 points) to obtain the spectrum at time $(n+1)\Delta t$ from the spectrum at time $n\Delta t$. 

\vspace{20ex}
\bibliography{bibliograph}
\end{document}